\documentstyle[11pt,amsmath,amssymb,epsf]{article}

\makeatletter
\@addtoreset{equation}{section}
\makeatother

\csname
@addtoreset\endcsname{equation}{section}

\def\charti{I}

\def\msl{\mathfrak{sl}}

\def\Real{{\mathbb R}}
\def\Comp{{\mathbb C}}

\def\1{1\hspace{-4pt}1}
\def\j1{\widetilde{1\hspace{-4pt}1}}

\def\bec{\begin{center}}
\def\ec{\end{center}}
\def\a{\alpha} \def\ad{{\dot{\a}}} 

\def\b{\beta} \def\bd{{\dot{\b}}} 
\def\c{\gamma} \def\cd{{\dot{\c}}}
\def\C{\Gamma}
\def\d{\delta} 

\def\e{\epsilon}

\def\F{\Phi}

\def\k{\kappa}
\def\l{\lambda}
\def\L{\Lambda}
\def\m{\mu}
\def\n{\nu}

\def\s{\sigma}
\def\S{\Sigma}
\def\t{\tau}

\def\O{\Omega}
\def\o{\omega}

\def\una{{\underline{\alpha}}}

\def\yb{{\bar y}}
\def\zb{{\bar z}}

\newcommand{\eq}[1]{(\ref{#1})}

\def\be{\begin{equation}}
\def\ee{\end{equation}}
\def\bea{\begin{eqnarray}}
\def\eea{\end{eqnarray}}
\def\ba{\begin{array}}
\def\ea{\end{array}}

\def\ad{\dot\alpha}
\def\bd{\dot\beta}

\def\bs#1\es{\begin{split}#1\end{split}}

\begin{document}

\begin{titlepage}

\setcounter{page}{1}

\begin{center}

\hfill MIFPA-11-8

\vskip 1cm

{\Large \bf  Geometry and Observables in Vasiliev's Higher Spin Gravity}

\vspace{12pt}

Ergin Sezgin\,$^1$ and Per Sundell\,$^{2,}$\footnote[3]{Ulysse Incentive Grant for Mobility in Scientific Research, F.R.S.-FNRS} \\

\vskip 25pt

{\em $^1$ \hskip -.1truecm George and Cynthia Woods Mitchell Institute for Fundamental Physics and Astronomy, Texas A\& M University, College Station,
TX 77843, USA}

{email: {\tt sezgin@tamu.edu}}

\vskip 15pt

{\em $^2$ \hskip -.1truecm Service de M\`ecanique et Gravitation, Universit\'e de Mons \\
20, Place du Parc, B7000 Mons, Belgium\vskip 5pt }

{email: {\tt per.sundell@umons.ac.de}} \\

\end{center}

\vskip 1cm

\begin{center} {\bf ABSTRACT}\\[3ex]
\end{center}

We provide global formulations of Vasiliev's four-dimensional minimal bosonic higher spin gravities by identifying structure groups, soldering one-forms and classical observables. In the unbroken phase, we examine how decorated Wilson loops collapse to zero-form charges and exploit them to enlarge the Vasiliev system with new interactions. We propose a metric phase whose characteristic  observables are minimal areas of higher spin metrics and on shell closed abelian forms of positive even degrees. We show that the four-form is an on shell deformation of the generalized Hamiltonian action recently proposed by Boulanger and one of the authors. In the metric phase, we also introduce tensorial coset coordinates and demonstrate how single derivatives with respect to coordinates of higher ranks factorize into multiple derivatives with respect to coordinates of lower ranks.

\end{titlepage}

\newpage

\tableofcontents

\pagebreak


\section{Introduction}


Vasiliev's higher spin gravities are fully nonlinear extensions of ordinary gravities by higher spin gauge fields as well as specific fields with spin less than two \cite{vasiliev,Vasiliev:2003ev} (for reviews, see \cite{Vas:star,Bekaert:2005vh}).
Their classical equations of motion admit anti-de Sitter or de Sitter spacetimes as exact solutions with unbroken higher spin symmetries. The corresponding fluctuation fields form higher spin multiplets consisting essentially of symmetric tensor gauge fields, known as Fronsdal tensors.
In this paper, we shall consider the four dimensional minimal bosonic models whose spectrum consist of a scalar, a metric and a tower of real Fronsdal tensors of even ranks greater than or equal to four each occurring once \cite{vasiliev,Vas:star,Us:analysis,Sezgin:2003pt}\footnote{The characterization of these models as minimal, essentially refers to the fact that their physical spectrum cannot be truncated further; for a more precise definition, see \cite{Us:analysis}.}

So far, Vasiliev's theory has been studied primarily at the level of the field equations and in terms of locally defined quantities in ordinary spacetimes.
It is clearly desirable to develop a globally defined framework for the classical theory, which is the primary aim of this paper.
This entails the identification of structure groups resulting in generalized metrics and other intrinsically defined observables.

The choice of structure group is not unique from kinematics alone.
Rather, it emulates various patterns of symmetry breaking leading to physically distinct global formulations of a given locally defined Vasiliev system which one may think of as representing different phases of the theory; for a related recent discussion, see also \cite{Colombo:2010fu}.

In this paper, we shall focus on four dimensional higher spin gravities for which we propose four natural choices for the structure group:
One choice is generated by the full higher spin algebra corresponding the unbroken phase with highly constrained observables, such as Wilson loops and zero-form charges \cite{Sezgin:2005pv,Colombo:2010fu}.
As the latter do not break any symmetries, they are on shell closed and can be used to introduce new interactions with free parameters into Vasiliev's fully nonlinear master equations.
%
%
A second choice for the structure group is the canonical $SL(2,\Comp)$ which is what has been assumed implicitly so far in the literature \cite{Vas:star}.
For remaining two choices, the structure algebras are infinite dimensional generalizations of ${\mathfrak{sl}}(2,\Comp)$.
One of these is projected out by particular, algebraically tractable automorphism already present in the local formulation and which is therefore more natural.
In the latter case, by making use of the generalized vielbein, we construct abelian $p$-forms and $p$-volume elements that can in principle be extremized.
The resulting $p$-form charges and extremal $p$-volumes are examples of intrinsically defined observables: functionals that are off shell invariant under generalized structure group gauge transformations, and on shell invariant modulo boundary conditions under the remaining Cartan gauge transformations. In particular, they are on shell invariant under diffeomorphisms.

Although the intrinsically defined observables are described in a coordinate-free fashion and at the fully nonlinear level, their evaluation on exact solutions may require the specification of, possibly infinitely many, tensorial coordinates corresponding to the generalized translations implied by the structure group.
As we shall illustrate below, the dependence of all perturbatively defined Weyl tensor components on the higher tensorial coordinates is completely constrained. For example, the derivative of the physical scalar with respect to the $n$-th rank symmetric traceless tensorial coordinate is expressed in terms of $n$th order derivative with respect to the four-dimensional spacetime coordinate.
The framework that we shall set up may also facilitate the nonlinear completions of earlier attempts at understanding higher spin geometry at the level of free fields \cite{de Wit:1979pe,Francia:2002aa}.


As for the $p$-form charges, they have a natural application in the off shell formulation of Vasiliev theory\footnote{For a related discussion of actions and charges in unfolded dynamics, see \cite{Vasiliev:2005zu}.}.
An intrinsically four dimensional action principle with Lagrangian proportional to constraints was proposed in \cite{Vasiliev:1988sa} and recently revisited in \cite{Doroud:2011xs}.
In an alternative formulation \cite{Boulanger:2011dd}, the four dimensional Vasiliev equations instead follow from a five dimensional action principle in which the four dimensional spacetime coincides with the boundary.
In addition, the latter formulation involves dual potentials and their Lagrange multipliers.
While these do not change the perturbative physical spectrum,
they can be used to add extra terms to the action, known as generalized Poisson structures, that
induce additional quantum corrections to the boundary correlation functions.
However, the resulting models remain topological in the sense that the original higher spin gauge symmetries remain unbroken and that the full action vanishes on shell (in the case of a single boundary).
Moreover, upon perturbative expansion, the boundary correlation functions do not contain the standard ones for Vasiliev theory including the two-and three-point functions of \cite{Giombi:2009wh,Giombi:2010vg}.

In this paper, as a step towards remedying the aforementioned problem, we propose deformations of the action by boundary functionals whose total variations vanish on shell, thus leaving the equations of motion unaltered and reducing on shell to intrinsically defined observables.
In particular, we construct such deformations that reduce to the aforementioned $p$-form charges and lead to additional boundary correlation functions.
Whether these are of the standard type for Vasiliev theory remains to be investigated.


The paper is organized as follows: we begin in the next Section by outlining certain key features of Vasiliev's formalism and presenting new interactions that we construct by using zero-form charges. Using these we then proceed to define global formulations in Section 3, stressing the role of the structure group. In Section 4, we examine the unbroken phase, where the structure group is the full higher spin algebra. In Section 5, we examine a broken phase of the theory, where the structure group is a certain natural extension of ${\mathfrak{sl}}(2,\Comp)$. In Section 6, we discuss the deformation of the action proposed in \cite{Boulanger:2011dd} by terms that are nonvanishing on shell. In Section 7, we summarize our results and comment on selected open problems. Appendix A contains details of the Lorentz covariantization. Appendix B describes a nonvanishing amplitude given by an integral over twistor space that depends on a two-form that arise in the duality extended formulation of the theory.


\section{Local Formulation}


The geometric formulation of four dimensional minimal bosonic higher spin gravity to be presented in the next Section rests on Vasiliev's unfolded master field formalism (see \cite{Vas:star, Bekaert:2005vh} for reviews).
In this Section we first present some of the general elements of this formalism and point out a generalization of the interaction function in the Vasiliev system. We shall then focus on the minimal bosonic models \cite{vasiliev,Us:analysis,Sezgin:2003pt} and cast their master equations into manifestly Lorentz-covariant form, referring to \cite{Vas:star} for the original proof of manifest local Lorentz invariance and \cite{Us:analysis} for more details.

\subsection{Vasiliev's Unfolded Master Field Formalism}

The Vasiliev equations for four-dimensional higher spin gravities are formulated using a set $(\widehat\Phi,\widehat A,\widehat J^{(r)})$, $r=1,\dots,N$, of Grassmann even differential forms of degrees $0,1$ and $2$, respectively, referred to as master fields. These fields are elements of the noncommutative and associative algebra
\be
\widehat {\cal A}~=~\O({\cal B})\otimes {\cal A}\ ,
\ee
where $\Omega({\cal B})$ is the space of differential forms on ${\cal B}$, a noncommutative base manifold, and ${\cal A}$ is a higher-spin algebra. In other words, the master fields are differential forms on ${\cal B}$ taking their values in ${\cal A}$.
The differential constraints read \cite{vasiliev}
\bea
\widehat F+\sum_r \widehat \eta^{(r)}(\widehat \Phi)\star  \widehat J^{(r)} &=& 0\ ,\label{master1}\\
\widehat D \widehat \Phi &=&  0\ ,\label{master2} \\
\widehat d \widehat J^{(r)} &=& 0\ ,\label{master3}
\eea
where the curvature and covariant derivative are defined by
\bea
\widehat F&=& \widehat d\,\widehat A+\widehat A\star \widehat A \ ,\qquad \widehat D\,\widehat \Phi\ =\ \widehat d\,\widehat \Phi
+\left[\widehat A,\widehat \Phi\right]_\pi\ ,
\eea
using the notation ($\widehat f,\widehat g\in\widehat{\cal A}$)
\be \left[\widehat f,\widehat g\right]_\pi~=~ \widehat f\star \widehat g-(-1)^{\widehat f\,\widehat g}\widehat g\star\pi(\widehat f)\ ,\ee
with $\pi$ denoting a (model dependent) automorphism of $\widehat{\cal A}$ obeying
\be \widehat d\pi~=~\pi\widehat d\ ,\qquad \pi^2~=~{\rm Id}\ .\ee
The algebraic constraints read
\be
\pi(\widehat J^{(r)})~=~\widehat J^{(r)}\ ,\qquad [\widehat f, \widehat J^{(r)}]_{\pi}~=~0 \ ,
\ee
for any element $\widehat f$ in the full algebra $\widehat{\cal A}$.
Finally, $\eta^{(r)}$ are $\star$-functions as follows:
\be
\widehat\eta^{(r)}(\widehat \Phi)~=~ \sum_{n=0}^\infty \eta_{2n+1}^{(r)}(\widehat\Phi) \,\widehat X^{\star n}\star \widehat \Phi\ ,\qquad \widehat X~:=~\widehat \Phi\star  \pi(\widehat \Phi)\ ,\label{calF}
\ee
where $\eta_{2n+1}^{{(r)}}(\widehat\Phi)$ are complex valued and on shell closed
functionals of $\widehat\Phi$, {\emph i.e.}
\be
\widehat d \,\eta_{2n+1}^{{(r)}}~=~0\ ,\qquad \left[\eta_{2n+1}^{{(r)}},\widehat f\,\right]_\star~=~0 \qquad \forall \widehat f\in\widehat{\cal A}\ .
\ee
In addition, the proper identification of Weyl tensors in the weak-field expansion constrains the constants $\eta^{(r)}_1(0)$. 
So far in the literature, $\eta_{2n+1}^{(r)}$ have been taken to be constants off shell. Here we propose to extend the  Vasiliev system by allowing $\eta_{2n+1}^{{(r)}}$ to depend on the master zero-form $\widehat\Phi$.
We shall come back to these deformations at the ends of Section \ref{sec:typeab} and Section \ref{sec:4}.

The definitions given so far suffice to show that the equations of motion are Cartan integrable, \emph{i.e.} compatible with $\widehat d^{\,2}\equiv 0$ on universal base manifolds, or equivalently, with the Bianchi identities $\widehat D\widehat F\equiv 0$ and $\widehat D^2 \widehat\Phi \equiv \left[\widehat F,\widehat \Phi\right]_\pi$. This is equivalent to that the twisted commutator $[\widehat \Phi,\eta^{(r)}(\widehat \Phi)]_\pi$ and the twisted covariant derivative $\widehat D\eta^{(r)}$ vanish modulo the equations of motion, as can be easily checked. It follows that the equations of motion remain invariant under the Cartan gauge transformations
\bea
\delta_{\widehat\e} \widehat A&=&\widehat D\widehat \e\ ,\qquad \delta_{\widehat\e}\widehat\Phi\ =\
-\left[\widehat\e,\widehat\Phi\right]_\pi\ ,\qquad \delta_{\widehat\e}\widehat J^{(r)}~=~0\ ,
\eea
with zero-form parameters $\widehat\e$. These transformations close to form the full higher spin algebra. The fact that $\widehat J^{(r)}$ remain invariant implies that they are globally defined elements in $\widehat{\cal A}$\,.

In order to apply the above formalism to four dimensional higher spin gravity, we need to specify further the local geometry of ${\cal B}$ and the attendant properties of $(\widehat J^{(r)},\pi)$. In this case, it is required that
\begin{itemize}

\item the symplectic leafs of ${\cal B}$ form a noncommutative twistor space ${\cal Z}$ of topology $\Comp^2$ with globally defined coordinates
\be Z^\una~=~(z^\a,\zb^{\ad})\ee
where $\una=1,\dots,4$ label an $Sp(4;\Real)$-quartet and $\a,\ad=1,2$ label two $SL(2;\Comp)$-doublets \cite{vasiliev};

\item the algebra ${\cal A}$ is a tensor product of the infinite-dimensional algebra $\Omega^{[0]}({\cal Y})$ of zero-forms on a copy ${\cal Y}$ of ${\cal Z}$ with coordinates $Y^\una=(y^\a,\yb^{\ad})$ and a finite-dimensional internal associative algebra ${\cal A}_{\rm int}$ \cite{Fradkin:1986ka,Konstein:1989ij}, \emph{viz.}
\be
{\cal A}~=~\O^{[0]}({\cal Y})\otimes {\cal A}_{\rm int}\ ,\label{calA}
\ee
associated with spacetime higher spin symmetries and internal Yang--Mills-like symmetries, respectively.
\end{itemize}
In what follows we assume that ${\cal B}$ has the bundle structure\footnote{We note that for certain exact solutions \cite{Didenko:2009td} the topology of the fiber space ${\cal Z}$ may vary as goes from one point to another on the base manifold in which case the bundle picture is replaced by a more sophisticated geometrical structure.}
\be
{\cal Z}~\hookrightarrow~ {\cal B}~{\longrightarrow}~{\underline{\cal M}}\ ,
\ee
where ${\underline{\cal M}}$ is a commuting manifold (or supermanifold in the case of supersymmetric models) and the projection amounts to restriction to some given base point in ${\cal Z}$, say $Z=0$.
The exterior derivative on ${\cal B}$ is thus given
\be
\widehat d~=~dX^{\underline M}\partial_{\underline M}+dz^\a\partial_\a +d\bar z^{\ad}\bar \partial_{\ad}\ ,
\ee
where $\partial_\a=\frac\partial{\partial z^\a}$, $\bar\partial_{\ad}=\frac\partial{\partial \bar z^{\ad}}$ and $X^{\underline M}$ coordinatize $\underline{\cal M}$. Without loss of generality, the twistor coordinates can be taken to obey the following canonical oscillator algebra
\bea [y_\a,y_\b]_\star&=&2i\e_{\a\b}\ ,\qquad [z_\a,z_\b]_\star\ =\ -2i\e_{\a\b}\ ,\\[5pt]
{}[\bar y_{\ad},\bar y_{\bd}]_\star&=&2i\e_{\ad\bd}\ ,\qquad  [\bar z_{\ad},\bar z_{\bd}]_\star\ =\ -2i\e_{\ad\bd}\ .\eea
and as the ${\cal Y}$ and ${\cal Z}$ spaces are two mutually commuting copies one has
\be [y_\a,z_\b]_\star\ =\ 0\ ,\qquad [\bar y_{\ad},\bar z_{\bd}]_\star\ =\ 0\ .\ee
One may also assume the reality conditions
\be (y^\a)^\dagger~=~\yb^{\ad}\ ,\qquad (z^\a)^\dagger~=~\zb^{\ad}\ ,\qquad \widehat d\circ \dagger~=~\dagger~\circ \widehat d\ .\ee
The algebra $\widehat {\cal A}=\Omega({\cal B})\otimes \Omega^{[0]}({\cal Y})\otimes {\cal A}_{\rm int}$ admits two automorphism $(\pi,\bar\pi)$ whose action on $\Omega({\cal B})\otimes \Omega^{[0]}({\cal Y})$ obeys $\widehat d(\pi,\bar\pi)=(\pi,\bar \pi)\widehat d$ and
\be \pi(X^{\underline M};y^\a,\yb^{\ad};z^\a,\zb^{\ad})~=~ (X^{\underline M};-y^\a, \yb^{\ad}; -z^\a, \zb^{\ad})\ ,\ee
\be \bar\pi(X^{\underline M};y^\a,\yb^{\ad};z^\a,\zb^{\ad})~=~ (X^{\underline M};y^\a, -\yb^{\ad}; z^\a, -\zb^{\ad})\ .\ee
Its action on $\Omega^{[0]}({\cal B})$ is inner and generated by the adjoint action by the Kleinian operators
\bea \widehat\kappa&=&\cos_\star (\pi \widehat N)\ ,\qquad \widehat{\bar\kappa}\ =\ \cos_\star (\pi \widehat {\bar N})\ ,\eea
where the $(\widehat N,\widehat{\bar N})$ are Weyl-ordered chiral number operators defined by
\be \widehat N\ =\ \frac12\left\{ \widehat a^-_\a,\widehat a^{+\a}\right\}_\star\ ,\qquad \left[\widehat a^-_\a,\widehat a^{+\b}\right]_\star~=~\delta_\a^\b\ ,\qquad \widehat {\bar N}~=~(\widehat N)^\dagger\ .\ee
Using the realization $(\widehat a^+_\a,\widehat a^-_\a)=\frac12( y_\a+z_\a,-i y_\a+i z_\a)$ yields
\be \widehat N~=~\frac{i}2 y^\a\star z_\a\ ,\qquad \widehat{\bar N}\ =\ -\frac{i}2 \bar y^{\ad}\star \bar z_{\ad}\ .\ee
The Kleinians can be expressed in various ordering schemes (for details, see \cite{Iazeolla:2008ix}); for the perturbative weak-field expansion it is convenient to normal order $(\widehat a^+_\a,\widehat a^-_\a)$ which yields \cite{vasiliev}
\be \widehat\kappa~=~\left[\exp(iy^\a z_\a)\right]_{\rm Normal}\ ,\qquad\widehat{\bar\k}~=~\left[\exp(-i\yb^{\ad}\zb_{\ad})\right]_{\rm Normal}\ ,
\ee
whereas for the purpose of performing chiral supertraces (see below) it is more convenient to work with Weyl order which yields\footnote{For the purpose of finding exact solutions it is more useful to factorize $\widehat\kappa=\kappa_y\star \kappa_z$ \cite{Didenko:2009td} where $\kappa_{y}$ and $\kappa_z$, respectively, are Kleinians for the chiral oscillators $y_\a$ and $z_\a$; one can then proceed by working in Weyl order where $\kappa_y=2\pi\delta^2(y)$ and $\kappa_z=2\pi\d^2(z)$ or normal orders where $\kappa_y$ and $\kappa_z$ are Gaussian. }
\be
\widehat\kappa~=~\left[(2\pi)^2 \delta^2(y)\delta^2(z)\right]_{\rm Weyl}\ ,\qquad \widehat{\bar\k}~=~\left[(2\pi)^2\delta^2(\yb)\delta^2(\zb)\right]_{\rm Weyl}\ .\label{Weyl}
\ee
Assuming in addition that the action of $(\pi,\bar \pi)$ on ${\cal A}_{\rm int}$ is inner and generated by two idempotent elements $(\Gamma_{\rm int},\bar \Gamma_{\rm int})$, it follows that there exists a doublet of globally defined two-forms $\widehat J^{(r)}=(\widehat J,\widehat{\bar J})$ given by
\be \widehat J ~=~  -\frac {i}4 dz^\a \wedge dz_\a \,\widehat\kappa\star \Gamma_{\rm int}\ ,\qquad \widehat{\bar J}~=~-\frac{i}4  d\bar z^{\ad}\wedge d\zb_{\ad} \, \widehat{\bar\kappa}\star\bar{\Gamma}_{\rm int}\ ,\ee
provided that the elements $\widehat f\in\widehat{\cal A}$ obey
\be \pi\bar\pi(\widehat f)~=~\widehat f\ .\ee
This condition assures that the model describes only bosonic degrees of freedom.

The key property for the choice made for $\widehat J^{(r)}$ is that the source term $\eta^{(r)}\star \widehat J^{(r)}=\eta\star \widehat J+\bar\eta\star \widehat{\bar J}$ cannot be redefined away. Moreover, it yields the correct linearized source terms for the two-form curvatures on ${\underline{\cal M}}$ upon reducing to $Z=0$ provided that $\eta_{1}(0)$ does not vanish \cite{vasiliev}.

\subsection{Minimal Bosonic Models and Type A and B Models}\label{sec:typeab}

In the case of bosonic models without Yang--Mills-like symmetries, one has ${\cal A}_{\rm int}=\mathbf 1$ and thus $\Gamma_{\rm int}=\bar \Gamma_{\rm int}=\mathbf 1$.
The master one-form is then given by
\be \widehat A~=~dX^{\underline M} \widehat A_{\underline M}(X,Y,Z)+dz^\a \widehat A_\a(X,Y,Z)+d\bar z^{\ad}\widehat A_{\ad}(X,Y,Z)\ ,\ee
and the master zero-form $\widehat \Phi=
\widehat \Phi(X,Y,Z)$. Bosonic models are obtained by imposing the conditions
\be \pi\bar\pi(\widehat A)~=~\widehat A\ ,\qquad \pi\bar\pi(\widehat \Phi)~=~\widehat \Phi\ ,\qquad \pi\bar\pi(\widehat J^{(r)})~=~\widehat J^{(r)}\ .\label{pibarpi}\ee
Minimal bosonic models whose weak-field expansions are in terms of symmetric tensors of even ranks require the stronger conditions
 \be \tau(\widehat A)~=~-\widehat A\ ,\qquad \tau(\widehat\Phi)\ =\ \pi(\widehat\Phi)\ ,\qquad \tau(\widehat J^{(r)})~=~-\widehat J^{(r)}\label{taucond}\ee
where the anti-automorphism $\tau$ is defined by
\be \t(X^{\underline M};y^\a,\yb^{\ad};z^\a,\zb^{\ad})~=~ (X^{\underline M};iy^\a,i \yb^{\ad}; -i z^\a,-i \zb^{\ad})\ .\ee
In the above, the $\tau$-conditions of the minimal bosonic model imply the $\pi\bar\pi$-conditions of the bosonic model.
Models with Lorentzian spacetime signature and negative cosmological constant require
\footnote{For other signatures, see \cite{Iazeolla:2007wt}.}
\be (\widehat A)^\dagger~=~-\widehat A\ ,\qquad (\widehat \Phi)^\dagger~=~\pi(\widehat \Phi)\ ,\label{daggercond2}\ee\be  (\widehat J)^\dagger~=~-\widehat {\bar J}\ ,\qquad (\eta(\l))^\dagger~=~\bar \eta(\l^\dagger)\ ,\label{daggercond}\ee
for any $\l\in\Comp$ and $\widehat J^{(r)}:=(\widehat J,\widehat{\bar J})$. Field redefinitions
\be \widehat \Phi~\rightarrow~ \widehat{\cal G}(\widehat \Phi)=g_1\, \widehat\Phi+g_3 \,\widehat\Phi\star\pi(\widehat\Phi)\star\widehat\Phi+\cdots\ ,\label{redef}\ee
where $g_{2n+1}(\widehat \Phi)$ are real valued and on shell closed
functionals of $\widehat\Phi$ , {\emph i.e.}

\be \widehat d g_{2n+1}~=~0\ ,\qquad (g_{2n+1})^\dagger~=~g_{2n+1}\ ,\qquad g_1(0)~\neq 0\ ,\ee
lead to the identifications
\be \widehat \eta(\widehat \Phi)~\sim~\widehat \eta(\widehat{\cal G}(\widehat \Phi))\ .\ee
These can be used to set
\be \widehat\eta(\widehat \Phi)~=~ e_\star^{i\widehat\Theta(\widehat X)}\star \widehat \Phi\ ,\label{phase}\ee
where $\widehat\Theta$ is a real $\star$-function and $\widehat X:=\widehat\Phi \star\pi(\widehat \Phi)$.

Assigning the scalar field intrinsic spacetime parity $\pm 1$, and imposing parity symmetry, drastically reduces the interaction freedom leading to the condition \cite{Sezgin:2003pt}
\be \widehat\eta~=~ \pm \widehat{\bar \eta}\ ,\ee
which fixes the phase factor in \eq{phase}, leading to the minimal bosonic
\be \mbox{Type A model}~:~ \widehat\eta~=~\widehat \Phi\ ,\ee
\be \mbox{Type B model}~:~ \widehat\eta~=~i\widehat \Phi\ .\ee
Thus, viewing the minimal model as a gauge theory on ${\underline{\cal M}}$, one has a gauge field $\widehat A_{\underline M}$ valued in the higher spin Lie algebra
\be \widehat{\mathfrak{hs}}(4)~=~\left\{\widehat P(Y,Z)~:~\tau(\widehat P)~=~(\widehat P)^\dagger~=~-\widehat P\right\}\ ,\qquad {\rm ad}_{\widehat P_1}(\widehat P_2)~=~\left[\widehat P_1,\widehat P_2\right]_\star\ ,\label{hs4}\ee
and coupled to algebraically constrained zero-forms $\widehat \Phi$ and $\widehat A_{\una}=(\widehat A_\a,\widehat A_{\ad})$ taking their values, respectively, in the twisted-adjoint representation
\be T[\widehat{\mathfrak{hs}}(4)]~=~\left\{\widehat T(Y,Z)~:~\tau(\widehat T)~=~(\widehat T)^\dagger~=~\pi(\widehat T)\right\}\ ,\qquad \rho(\widehat P)(\widehat T)~=~[\widehat P,\widehat T]_\pi\ ,\label{ths4}\ee
and the quasi-adjoint representation
\bea T'[\widehat{\mathfrak{hs}}(4)]~&=&~\left\{\widehat T'_{\una}(Y,Z)~:~\tau(\widehat T'_{\una})~=~-i\widehat T'_{\una}\ ,\quad (\widehat T'_{\a})^\dagger~=~\widehat T'_{\ad}\ \right\}\ ,\\[5pt] \rho'(\widehat P)(\widehat T'_{\una})~&=&~[\widehat P,\widehat T'_{\una}]_\star\ ,\eea
in accordance with \eq{taucond} and \eq{daggercond2}.


\subsection{Manifest Lorentz Covariance}\label{sec:lor}


As was shown in \cite{Vas:star} and studied further in \cite{Us:analysis}, manifest local Lorentz invariance can be achieved by first introducing the deformed oscillators
\bea \widehat S_\a&=& z_\a-2i\widehat A_\a\ ,\qquad \widehat S_{\ad}\ =\ \bar z_{\ad}-2i \widehat A_{\ad}\ .\label{Sa}\eea
In terms of these master fields, the twistor space equations take the form of a generalized version of Wigner's deformed oscillator algebra \cite{vasiliev}\footnote{Wigner's deformed oscillator algebra reads
$[S,S^\dagger]_\star = 1+\nu (-1)^N$ where $\nu\in \Real$ and $S=S(a,\a^\dagger)$ with $[a,a^\dagger]_\star=1$ and $N=a^\dagger\star a$. This algebra can be extended first into $[S_i,S^{\dagger,j}]=\delta_i^j+\nu\,(-1)^N\nu\pi_i^j$ where $i=1,2$ and $S_i=S_i(a_j,a^{\dagger j})$ with $[a_i,a^{\dagger j}]=\delta_i^j$ and $N=a^{\dagger i}\star a_i$. Vasiliev's extension then consists of complexifying the oscillators and replacing the ``anyon statistics'' parameter $\nu$ by a function of the oscillators.}:
\be \widehat S_{[\a}\star\widehat S_{\b]}\ =\  -i\e_{\a\b}(1-\eta\star\widehat\kappa\star \gamma_{\rm int})\ ,\qquad
\widehat S_{[\ad}\star\widehat S_{\bd]}\ =\  -i\e_{\ad\bd}(1-\bar \eta\star\widehat{\bar\kappa}\star \bar\gamma_{\rm int})\ ,\label{wigner1}\ee\\[-40pt]\be \widehat S_\a\star\widehat S_{\ad}\ =\ \widehat
S_{\ad}\star\widehat S_\a\ ,\label{wigner2}\ee\\[-40pt]\be \widehat S_\a\star\widehat\Phi+\widehat\Phi\star \pi(\widehat S_\a)\ =\ 0\ ,\qquad
\widehat S_{\ad}\star\widehat\Phi+\widehat\Phi\star \bar\pi(\widehat S_{\ad})\ =\ 0\ .\label{wigner3}\ee
The canonical Lorentz transformations of $(\widehat\Phi,\widehat S_\a,\widehat S_{\ad})$ are generated by the twisted-adjoint and adjoint actions of
\be
\widehat M_{\a\b}~=~ \widehat M^{(0)}_{\a\b}+\widehat M^{(S)}_{\a\b} \ ,\qquad \widehat M_{\ad\bd}\ =\ \widehat
M^{(0)}_{\ad\bd}+\widehat M^{(S)}_{\ad\bd}\ ,\ee\be  \widehat M^{(0)}_{\a\b}~=~y_{(\a}\star  y_{\b)}-z_{(\a}\star
z_{\b)}\ ,\qquad \widehat M^{(0)}_{\ad\bd}\ =\ \yb_{(\ad}\star
\yb_{\bd)}-\zb_{(\ad}\star \zb_{\bd)}\ ,\ee\be \widehat M^{(S)}_{\a\b}~=~
 \widehat S_{(\a}\star
\widehat S_{\b)}\ ,\qquad \widehat M^{(S)}_{\ad\bd}~=~ \widehat S_{(\ad}\star \widehat
S_{\bd)}\ .\ee
Thus, letting $(\omega^{\a\b}_{\underline M},\omega^{\ad\bd}_{\underline M})$ be an ${\mathfrak sl}(2,\Comp)$-valued connection and shifting \cite{Vas:star,Us:analysis}
\be \widehat A_{\underline M}~=~ \widehat W_{\underline M}+ {1\over 4i}\left(\omega_
{\underline M}^{\a\b}\widehat
M_{\a\b}+\bar\omega_{\underline M}^{\ad\bd}\widehat M_{\ad\bd}\right)\ ,\ee
the constraints on ${\underline{\cal M}}$  takes
the following manifestly Lorentz covariant form (see Appendix \ref{App:B}):
\be \nabla \widehat W+\widehat W^2+\widehat r^{(0)}+\widehat r^{(S)}~=~0\ ,\label{cW}\ee\be
\nabla \widehat \Phi+[\widehat W,\widehat \Phi]_\pi~=~0\ ,\qquad
\nabla \widehat S_\a+[\widehat W,\widehat S_\a]_\star~=~0\ ,\qquad
\nabla \widehat S_{\ad}+[\widehat W,\widehat S_{\ad}]_\star~=~0\
,\label{cS}\ee
where we have defined
\be \widehat r^{(0)}~=~{1\over 4i}\left(r^{\a\b}\widehat
M^{(0)}_{\a\b}+\bar r^{\ad\bd}\widehat
M^{(0)}_{\ad\bd}\right)\ ,\qquad \widehat r^{(S)}~=~{1\over 4i}\left(r^{\a\b}\widehat
M^{(S)}_{\a\b}+\bar r^{\ad\bd}\widehat
M^{(S)}_{\ad\bd}\right)\ ,\label{widehatr}\ee
\be r^{\a\b}~=~d\omega^{\a\b}-\omega^{\a\c}\wedge
\omega_{\c}{}^{\b}\ ,\qquad \bar r^{\ad\bd}~=~(r^{\a\b})^\dagger~=~d\bar\omega^{\ad\bd}-\bar\omega^{\ad\cd}\wedge
\bar \omega_{\cd}{}^{\bd}\ ,\label{Riemann}\ee
and the Lorentz covariant derivative
\be \nabla~=~d+\rho(\o)\label{nabla}\ee
act in the canonical representation $\rho(\L)\equiv \delta_\L$ of ${\mathfrak sl}(2,\Comp)$ given by
\be\delta_\L \widehat W~=~ -\left[\widehat\L^{(0)},\widehat W\right]_\star\
,\qquad \delta_\L\widehat\o^{(0)}~=~d\widehat\L^{(0)}+\left[\widehat\o^{(0)},\widehat\L^{(0)}\right]_\star\ ,
\ee\be\delta_\L\widehat \Phi~=~ -\left[\widehat \L^{(0)},\widehat \Phi\right]_\star\ ,\qquad\delta_\L\widehat S_\a~=~ \Lambda_\a{}^\b\widehat S_\b-\left[\widehat \L^{(0)},
\widehat S_\a\right]_\star\ ,\label{LT1}\ee
where the ${\mathfrak sl}(2,\Comp)$-valued gauge field and gauge parameter are given by
\be\widehat\o^{(0)}~=~{1\over 4i}\left(\o^{\a\b}\widehat
M^{(0)}_{\a\b}+\overline{ \o}^{\ad\bd}\widehat
M^{(0)}_{\ad\bd}\right)\ ,\qquad \widehat\L^{(0)}~=~{1\over 4i}\left(\Lambda^{\a\b}\widehat
M^{(0)}_{\a\b}+\overline{ \Lambda}^{\ad\bd}\widehat
M^{(0)}_{\ad\bd}\right)\ ,\ee
and the Lorentz transformation $\delta_\L\widehat\o^{(0)}$ amounts to
\be \delta_\L \omega^{\a\b}~=~ d\Lambda^{\a\b}-2\o^{\c(\a}\L_{\c}{}^{\b)}\ .\label{LT2}\ee
The Lorentz-covariantized Cartan gauge transformations take the form
\be \delta_{\widehat\e}\,\widehat W ~=~ \nabla\,\widehat \e+[\widehat W,\widehat \e\,]_\star\ ,\qquad\delta_{\widehat\e}\,\widehat \o^{(0)}~=~0\ ,\ee\be  \delta_{\widehat\e}\widehat \Phi~=~-[\widehat\epsilon,\widehat\Phi]_\pi\ ,\qquad \delta_{\widehat\e}\,\widehat S_\a~=~ -[\widehat \e,\widehat S_\a]_\star\ .\ee
The introduction of the canonical Lorentz connection leads to an over-parameterization of $\widehat A_{\underline M}$, thus inducing the shift symmetry
\be \delta_\eta \widehat W_{\underline M}~=~-\frac1{4i}\left(\eta_{\underline M}^{\a\b}\widehat M_{\a\b}+\bar\eta_{\underline M}^{\ad\bd}\widehat M_{\ad\bd}\right)\ ,\qquad  \delta_\eta (\o^{\a\b}_M,\bar\o^{\ad\bd}_M)~=~ (\eta^{\a\b}_{\underline M},\bar\eta^{\ad\bd}_{\underline M})\ ,\ee
\be \delta_\eta \widehat\Phi~=~0\ ,\qquad \delta_\eta \widehat S_\a~=~0\ ,\label{shiftsymm}\ee
with unconstrained parameters $(\eta_{\underline M}^{ab},\bar\eta_{\underline M}^{\ad\bd})$ and acting such that
$\delta_{\eta} \widehat A_{\underline M}=0$, which ensures shift symmetry of the master equations. Provided that $\widehat M_{\a\b}$ has nontrivial components along $y_{(\a}\star y_{\b)}$\,, this symmetry can be used to impose the condition that $\widehat W$ has vanishing components along $y_{(\a}\star y_{\b)}$\,.

The Vasiliev equations can thus be written on two equivalent forms, namely
\begin{itemize}
\item as in Eqs. \eq{master1}--\eq{master3}, which one may refer to as an associative quasi-free differential algebra on an extended correspondence space\footnote{This terminology refers to the fact that one has a parent theory formulated on a higher-dimensional space that contains projections to theories formulated on lower-dimensional submanifolds. These latter two theories thus correspond to each other via the parent theory and one may refer to the space in which the latter is defined as the correspondence space. In this sense and taking into account dynamical symmetry breaking mechanisms, one may think of Vasiliev's formulation of higher spin gravity as the natural framework for understanding the correspondence between the formulations of gauge theory in four-dimensional spacetime and in Penrose's twistor space.}, namely the product space ${\cal B}\times {\cal Y}$; or
\item as in Eqs. \eq{wigner1}--\eq{wigner3}, \eq{cW} and \eq{cS}, which one may refer to as a graded commutative quasi-free differential algebra on a commutative base manifold ${\underline{\cal M}}$ with fiber generated by functions on ${\cal Y}\times {\cal Z}$, generalized curvature constraints \eq{cW} and \eq{cS} and algebraic zero-form constraints \eq{wigner1}--\eq{wigner3};
\end{itemize}
The correspondence space formulation in the spirit of string field theory, while the quasi-free differential algebra formulation relies on gauging a higher spin Lie algebra on an ordinary commutative base manifold.


\section{Global Formulation}\label{sec:global}

In this section we set up the global formulation of Vasiliev's minimal bosonic models based on the notions of structure groups, intrinsically defined observables and other geometric data; for related discussions,  see \cite{Boulanger:2008up,Colombo:2010fu,Boulanger:2011dd}.

\subsection{Structure Groups and Observables}

In order to provide a formulation of Vasiliev's theory that is defined globally on ${\underline{\cal M}}$ a structure group needs to be defined. By its definition, the structure group is generated by a structure algebra given by a subalgebra %
\be \widehat{\mathfrak t}~\subseteq~\widehat{\mathfrak{hs}}(4)\oplus {\mathfrak{sl}}(2,\Comp)\ .\ee
Decomposing the manifold ${\underline{\cal M}}$ into coordinate charts ${\underline{\cal M}}_I$ labelled by $I$, \emph{i.e.}
\be {\underline{\cal M}}~=~\bigcup_I {\underline{\cal M}}_I\ ,\ee
the classical moduli space then consists of gauge orbits of locally defined field configurations \be \left\{{\widehat W}_\charti,\o_I\,,\,\widehat\F_\charti\, ,\,\widehat S_{I\a}\,,\, \widehat S_{I\ad}\right\}\label{config}\ee
glued together by transition functions $\widehat G_
{\charti}^{\charti'}=\exp_\star(\widehat t_{\charti}^{\charti'})$ with $\widehat t_{\charti}^{\charti'}\in \widehat{\mathfrak t}$ defined on ${\underline{\cal M}}_I\cap {\underline{\cal M}}_{I'}$.
The classical moduli space thus encodes a principal ${\widehat{\mathfrak t}}$-bundle with connection
\be \widehat \C ~=~\Pi_{\widehat{\mathfrak t}}(\widehat W\oplus \o)\ ,\ee
where $\Pi_{\widehat{\mathfrak t}}$ denotes the projection to ${\widehat{\mathfrak t}}$, and an associated ${\widehat{\mathfrak t}}$-bundle with section $(\widehat E,\widehat \Phi,\widehat S_\una)$
where
\be \widehat E~=~(1-\Pi_{\widehat{\mathfrak t}})(\widehat W\oplus \o)\ .\ee
The classical observables are thus functionals  ${\cal O}\left[{\widehat W}_\charti,\o_I,\widehat\F_\charti ,\widehat S_{I\a},\widehat S_{I\ad}\right]$ whose invariance properties under gauge transformations with locally defined parameters $\left\{\widehat \e_I\oplus \L_I\right\}$ are:
\begin{itemize}
\item $\delta_{\widehat\L_I}{\cal O}\equiv 0$ off shell for locally defined $\widehat{\mathfrak{t}}$-valued parameters $\widehat \L_I=\Pi_{\widehat{\mathfrak t}}(\widehat\e_I\oplus {\L_I})$ (defined independently for each $I$); and
\item $\delta_{\{\widehat\xi_I\}}{\cal O}= 0$ on shell for parameters $\widehat \xi_I=(1-\Pi_{\widehat{\mathfrak t}})(\widehat\e_I\oplus \L_I)$ that form sections of a ${\widehat{\mathfrak t}}$-bundle associated to the principal bundle.
\end{itemize}
Taken together, these two conditions imply that the observables are left invariant on shell by the diffeomorphisms of ${\underline{\cal M}}$.

Each structure group corresponds to a specific phase of the theory with its own classical moduli space and associated set of classical observables.
The minimal bosonic higher spin gravity has four natural phases:
\begin{itemize}
\item an unbroken phase with structure algebra $\widehat{\mathfrak{hs}}(4)\oplus \msl(2,\Comp)$ and classical observables given by decorated Wilson lines and zero-form charges, which we shall treat in more detail in Section \ref{sec:4};
\item a broken phase with $\pi$-even higher spin structure algebra $\widehat{\mathfrak {hs}}_+(4)\oplus \msl(2,\Comp)$ where
\be \widehat{\mathfrak {hs}}_+(4)~=~\frac12(1+\pi)\widehat{\mathfrak{hs}}(4)\ ,\label{hsplus}\ee
and characteristic classical observables given by minimal areas and charges of on shell closed abelian forms, which will be treated in Section \ref{sec:5};
\item a broken phase with chiral higher spin structure algebra\footnote{This choice was pointed out to us by M. Vasiliev.} $\widehat{\mathfrak {hl}}(2,\Comp)\oplus \msl(2,\Comp)$ where
\be \widehat{\mathfrak {hl}}(2,\Comp) ~=~\left\{ \widehat P(y,z)+\widehat {\bar P}(\yb,\zb)\right\} \cap \widehat{\mathfrak{hs}}(4)\ ,\ee
that is, all polynomials in $\widehat{\mathfrak {hs}}_+(4)$ that are purely holomorphic or anti-holomorphic. The geometrical significance of this choice of structure group and its consequences for deforming the theory off shell remains to be investigated;
\item the broken Lorentz invariant phase with structure algebra given by the canonical ${\mathfrak{sl}}(2,\Comp)$ which is implicitly assumed in most of the existing literature on higher spin gravity.

\end{itemize}

\subsection{Initial Data, Transition Functions and Boundary Values}\label{sec:data}

Two locally defined configurations \eq{config} belong to the same gauge orbit if they are related by locally defined gauge transformations $\widehat G_I=\exp_\star\widehat t_I$ where $\widehat t_I$ are $\widehat{\mathfrak{t}}$-valued otherwise unrestricted functions on ${\underline{\cal M}}_I$. The locally defined configurations can thus be written as \cite{Bolotin:1999fa,Sezgin:2005pv}\footnote{For a more general discussion of gauge functions and initial data in Cartan
integrable systems, see \cite{Boulanger:2011dd}.}
\bea \widehat W_I&=& (\widehat L_I)^{-1}\star d\widehat L_I- {1\over 4i}\left(\omega_
{I \underline M}^{\a\b}\widehat
M_{\a\b}+\bar\omega_{I \underline M}^{\ad\bd}\widehat M_{\ad\bd}\right)\ ,\label{wi}\\[2pt]
\widehat \Phi_I&=&(\widehat L_I)^{-1}\star \widehat \Phi'_I(Y,Z)\star \pi(\widehat L_I)\ ,\\[2pt]
\widehat S_{I\una}&=&(\widehat L_I)^{-1}\star \widehat S'_{I\una}(Y,Z)\star \widehat L_I\ ,\eea
where $\widehat L_I$ are gauge functions valued in the coset generated by $\widehat{\mathfrak{hs}}(4)/\widehat{\mathfrak t}$. The gauge functions and initial data are related across chart boundaries via transition functions $\widehat G_I^{I'}$ and constant monodromy matrices\footnote{In going from one chart to another, one has $d(\widehat L_{I'} \star \widehat G_{I}^{I'} \star \widehat L_{I'}^{-1})=0$, which implies that $\widehat L_{I}\star \widehat  G_{I'}^{I}  \star \widehat L_{I'}^{-1}=:\widehat M_{I'}^{I}$, a set of constant group elements defining a representation of the first homotopy group of the base manifold, referred to as the monodromy group.} $\widehat M_I^{\prime\, I'}$ as follows:
\bea \widehat L_I&=&(\widehat M_{I}^{\prime\,I'})^{-1}\star \widehat L_{I'}\star \widehat G_I^{I'}\ ,\\[2pt] \widehat \Phi'_{I}&=& (\widehat M_{I}^{\prime\,I'})^{-1}\star \widehat \Phi'_{I'}\star \widehat M_I^{I'}\ ,\\[2pt]
\widehat S^{\,\prime}_{I\una }&=& (\widehat M_{I}^{\prime\,I'})^{-1}\star S^{\,\prime}_{I'\una }\star \widehat M_I^{I'}\ .
\eea
Assuming that $\left.\widehat L_I\right|_{p_I}=\mathbf 1$
for base points $p_I\in {\underline{\cal M}}_I$, the reduced master fields
\be \widehat \Phi'_I~=~\left.\widehat \Phi_I\right|_{p_I}\ ,\qquad \widehat S^{\,\prime}_{I\una}(Y,Z)~=~\left.\widehat S_{I\una}\right|_{p_I}\ ,\ee
obey the deformed oscillator algebra \eq{wigner1}--\eq{wigner3} to be solved subject to initial data
\be C_I(Y)~=~\left.\widehat \Phi'_I\right|_{Z=0}\ ,\ee
and boundary conditions on $\widehat S^{\,\prime}_{I\una}$ that we refer to as $S$-moduli (the latter are related to projectors in the $\star$-product algebra; for example, see \cite{Iazeolla:2007wt}). Thus, globally defined solutions can be constructed from gauge functions, transition functions, monodromies and the initial data in single chart, say at $p_{I_0}\in {\underline{\cal M}}_{I_0}$. The resulting classical moduli space is thus generated from classical observables depending on (see also \cite{Boulanger:2011dd})
\begin{itemize}
\item invariants for the initial data at $p_0$ in the form of $C_{I_0}(Y)$ and $S_{I_0}$-moduli where $I_0$ labels a chart containing $p_0$;
\item the homotopy classes $\left[\widehat G_I^{I'}\right]=\left\{\widehat G^{-1}_{I'}\star \widehat G_I^{I'}\star \widehat G_{I}\right\}$ and monodromies $\widehat M^{\prime I}_I$ ;
\item the gauge equivalence classes of boundary gauge functions $\widehat L_I|_{\partial{{\underline{\cal M}}}}$.
\end{itemize}

The classical observables depend on the gauge functions via their values at $\partial{\underline{\cal M}}$ modulo boundary gauge transformations.
In summary, the unfolded equations of motion on ${\underline{\cal M}}_I$ can be integrated using gauge functions and initial data given by the values of the zero-forms at a single point in ${\underline{\cal M}}_I$, which implies that no new strictly local degrees of freedom are introduced if new dimensions are added to ${\underline{\cal M}}$. When combined with the soldering mechanism this can be used to formulate higher spin gravity in extended spacetimes with higher tensorial coordinates as we shall exemplify in Section \ref{sec:5}.


\section{Unbroken Phase: Wilson Loops and Zero-Form Charges}\label{sec:4}

In the unbroken phase, the principal bundle has the connection
\be \widehat \C~=~\widehat W\oplus {\o}\ ,\ee
and all Cartan gauge symmetries as well as local Lorentz symmetries remain locally defined. The canonical Lorentz connection can be shifted away by choosing the gauge
\be \o~=~\bar\o~=~0\quad \Rightarrow\quad\widehat \Gamma~=~\widehat W\ .\ee
As a result, the Vasiliev equations consist of the deformed oscillator algebra \eq{wigner1}--\eq{wigner3} and the differential constraints
\be d\widehat W+\widehat W\star \widehat W~=~0\ ,\qquad d\widehat \Phi+[\widehat W,\widehat\Phi]_\pi~=~0\ ,\qquad d\widehat S_{\una}+[\widehat W,\widehat S_\una]_\star~=~0\ ,\label{dw}\ee
where $d=dX^{\underline M}\partial_{\underline M}$. To define classical observables as functionals of the master fields that do not depend on the ordering prescription, a trace operation on the oscillator algebra is needed. We shall use
\bea \widehat {\rm Tr}\left[\widehat f(y,z,dz;\bar y,\bar z,d\bar z)\right]&=& \int_{\cal Y} \frac{d^2y }{2\pi} \frac{d^2\bar y }{2\pi} \int_{\cal Z} \widehat f(y,z,dz;\bar y,\bar z,d\bar z)\ ,
\label{Tr}\eea
with normalization and integration domain chosen such that
\be  \widehat {\rm Tr}'\left[\widehat f(y,z,\bar y,\bar z)\right]~:=~\widehat{\rm Tr}\left[ \frac{d^2z }{2\pi} \frac{d^2\bar z }{2\pi}  \widehat f (y,z,\bar y,\bar z)\right]~=~ \int_{\cal Y} \frac{d^2y }{2\pi} \frac{d^2\bar y }{2\pi} \int_{\cal Z} \frac{d^2z }{2\pi} \frac{d^2\bar z }{2\pi} \widehat f(y,z;\bar y,\bar z)\ ,
 \label{Trprime}\ee
where $(y,z)$ and $(\bar y,\bar z)$ are treated as real and independent variables, and we are using the convention that the integration operation $\int_{\cal Z}(\cdot)$ projects onto the top form degree in ${\cal Z}$.
Traces given by convergent integrals are expected to obey the following key properties:
\begin{itemize}
\item independence of the choice of ordering scheme used to represent operators by symbols;
\item the cyclicicity property;
\item $\pi$-invariance, \emph{i.e.}
\be \widehat{\rm Tr}[\pi(\widehat f)]~=~\widehat{\rm Tr}[\widehat f]\ ,\ee
\emph{idem} $\bar\pi$, and for this reason we refer to $\widehat{\rm Tr}$ as the chiral trace.
\end{itemize}
The insertion of inner Kleinians into the chiral trace operation yields graded cyclic chiral trace operations $\widehat{\rm Tr}\left[\widehat\kappa\star (\cdot)\right]$, $\widehat{\rm Tr}\left[\widehat{\bar\kappa}\star (\cdot)\right]$ and $\widehat{\rm Tr}\left[\widehat\kappa\widehat{\bar\kappa}\star (\cdot)\right]$ referred to as chiral supertraces of the oscillator algebra. The last one reduces to a cyclic trace operation in the bosonic model due to \eq{pibarpi}. The Kleinian $\widehat \k$  produces phase factors that localize the chiral integral to $y=z=0$ in the Weyl ordering scheme as can be seen from \eq{Weyl}, and \emph{idem} $\widehat{\bar \kappa}$.

Natural classical observables in the unbroken phase are Wilson loops along paths $\gamma\subset {\underline{\cal M}}$ with impurities in the form of adjoint vertex operators inserted at points $x_i\in \gamma$ for $i=1,\dots,N$. A basis for these observables consists of ($\s=0,1$)
\bea {\cal W}^\s_{\gamma;\{x_i\}}\left(k_i,\bar k_i;\l_i,\bar \l_i\right)&=& \widehat {\rm Tr}'\left[(\widehat\kappa\widehat{\bar\kappa})^\s \star{\rm P} \left\{\prod_{i=1}^N \left.{\cal V}^{k_i,\bar k_i}_{\l_i,\bar \l_i}(\widehat \Psi,\widehat{\bar\Psi},\widehat S)\right|_{x_i}\star
U_\gamma[\widehat W]\right\}\right]\ ,\eea
where we have defined
\begin{itemize}\item the vertex operator \be {\cal V}^{k,\bar k}_{\l,\bar \l}(\widehat \Psi,\widehat{\bar\Psi},\widehat S)~=~\exp_\star \left[i( \l^\a\widehat S_{\a}+\bar\l^{\ad}\widehat S_{\ad})\right]\star
(\widehat \Psi)^{\star k}\star (\widehat{\bar\Psi})^{\star \bar k}\ ,\label{vertex}\ee
with $k,\bar k\in\mathbb N$\,, $(\l^\a,\bar\l^{\ad})$ being auxiliary twistor variables and the adjoint elements
\be \widehat \Psi~=~\widehat \Phi\star\widehat\kappa\ ,\qquad \widehat{\bar\Psi}~=~\widehat \Phi\star \widehat{\bar\kappa}\ ,\ee
obeying the following relations
\be d\widehat \Psi+[\widehat W,\widehat\Psi]_\pi~=~0\ ,\qquad d\widehat {\bar \Psi}+[\widehat W,\widehat {\bar \Psi}]_\star~=~0\ ,\label{dpsi}\ee
\be \{\widehat S_\a,\widehat \Psi\}_\star~=~[\widehat S_\a,\widehat {\bar \Psi}]_\star~=~0\ ,\qquad \{\widehat S_{\ad},\widehat {\bar\Psi}\}_\star~=~[\widehat S_{\ad},\widehat \Psi]_\star~=~0\ ,\label{sss2}\ee
\be \widehat \Psi\star \widehat \Psi ~=~\widehat{\bar \Psi}\star \widehat{\bar \Psi}\ ,\qquad [\widehat\Psi,\widehat{\bar\Psi}]_\star~=~0\ ,\qquad \widehat \Psi\star \widehat\kappa\widehat{\bar\kappa}~=~\widehat{\bar\Psi}\ ;\label{psipsi}\ee
\item the path-ordered exponential $U_\gamma[\widehat W]$ is defined by
\be {\rm P}~\left\{{\cal V}\star U_\gamma[\widehat W]\right\}~:=~{\rm P}\left\{{\cal V}\star \prod_l \left.\left(\widehat G_{I_{l-1}}^{I_l}\right)\right|_{p_l}\star \exp \int_{\c_l} \widehat W_{I_l}\right\}\ ,\ee
where $\gamma$ has been cut into oriented links $\gamma_l$ with end-points $p_l$ and $p_{l+1}$ passing through charts $I_l$, \emph{viz.}
\be \gamma~=~\cup_l \gamma_l\ ,\qquad \partial\gamma_l~=~\{p_{l+1}\}-\{p_l\}\ ,\ee
where the sign indicates the orientation, and insertions of transition functions $\widehat G_{I_{l-1}}^{I_l}$ in passages from $\gamma_{l-1}$ to $\gamma_l$ assure gauge invariance.
\end{itemize}
From \eq{wi} and since $\o=\bar\o=0$ it follows that if there is no decoration along a link $\gamma_l$ then
\be {\rm P}\exp \int_{\c_l} \widehat W_{I_l}~=~ \left.(\widehat L_{I_l})^{-1} \right|_{p_l}\star \left.(\widehat L_{I_l})\right|_{p_{l+1}}\ .\ee
Thus, if ${\cal V}=1$ and assuming trivial monodromies, then \be {\rm P} \left\{U_\gamma[\widehat W]\right\}~=~\mathbf{1}\ .\label{1}\ee
Formally, the flatness conditions \eq{dw} and \eq{dpsi} imply that the Wilson loops remain invariant under smooth deformations of $\c$ and $x_i$. If it is possible to
move all impurities to a single point on $\gamma$, say $x$, then
\bea {\cal W}^\s_{\c;x}\left(k,\bar k;\l,\bar\l\right)&=&\widehat {\rm Tr}'\left[ (\widehat\kappa\widehat{\bar\kappa})^\s\star{\rm P}  \left\{\left.{\cal V}^{k,\bar k}_{\l,\bar\l}(\widehat \Psi,\widehat{\bar\Psi},\widehat S)\right|_{x}\star
U_\gamma[\widehat W]\right\}\right]\ ,\eea
which can be collapsed further using \eq{1}. Thus, provided no singularities appear, the decorated Wilson loops are equivalent to the zero-form charges ($\s=0,1$)
\bea {\cal I}^\s\left(k,\bar k; \l,\bar\l\right)~=~\widehat{\rm Tr}'\left[(\widehat\kappa\widehat{\bar{\kappa}})^\s\star {\cal V}^{k,\bar k}_{\l,\bar\l}(\widehat \Psi,\widehat{\bar\Psi},\widehat S)\right]\ .\eea
Whether the vertex operator \eq{vertex} is well-defined and the zero-form charge is finite depends on the details of the twistor-space behavior of the master fields, which are determined on-shell by the choice of boundary conditions in twistor space. 
For example, one may require all master fields to be real analytic in $Y$ and $Z$ (for generic $X$) and impose $\widehat \Phi\vert_{Z=0}=\Phi:(X;Y)$
and $Z^{\underline \a} \widehat A_{\underline\a}=0$.
One may then proceed by expanding the master fields perturbatively in the initial datum $\Phi$ (for the resulting generally covariant weak-field expansion of the equations of motion in $X$ space, see for example \cite{Us:analysis}).
In this scheme, it makes sense to expand the vertex operator, including the exponential, in $\Phi$, retaining, however, the zeroth order factor $\exp_\star \left[i\left(\l^\a z_\a+\bar\l^{\ad}_{\ad}\right)\right]$ as to dampen the $Z$-space integral in the zero-form charge. 
As for the $Y$-space integral, its finiteness properties are related the choice of functional class for $\Phi$ in $Y$-space, which corresponds to choosing higher-spins representations for the linearized theory 

In an alternative approach, one may seek to evaluate the zero-form charge by expanding the vertex operator inside the trace in $\l$ and $\bar\l$, which leads to expressions of the form
\bea \widehat{\rm Tr}'\left[(\widehat\kappa\widehat{\bar{\kappa}})^\s\star \widehat S_{(\a_1}\star\cdots \star\widehat S_{\a_n)}\star \widehat S_{(\ad_1}\star\cdots \star\widehat S_{\ad_{\bar n})}\star
(\widehat \Psi)^{\star k}\star (\widehat{\bar\Psi})^{\star \bar k}\right]\ .\eea
If $n+\bar{n}>0$ then it follows from \eq{sss2} and \eq{sss} that the argument of the trace is a sum of commutators of polynomials in deformed oscillators. 
Whether these commutators vanish depend on the convergence properties of the separate terms in these commutators, \emph{i.e.} on the fall-off behavior of the deformed oscillators and zero-forms at the boundaries of twistor space. 

Setting $n+\bar{n}=0$ yields the Lorentz-scalar zero-form charges \cite{Sezgin:2005pv,Iazeolla:2007wt}
\bea {\cal I}^\s\left(k,\bar k\right)&=&\left.{\cal I}^\s\left(k,\bar k; \l,\bar\l\right)\right|_{(\l,\bar\l)=(0,0)}~=~\widehat {\rm Tr}'\left[(\widehat\kappa\widehat{\bar{\kappa}})^\s\star(\widehat \Psi)^{\star k}\star (\widehat{\bar\Psi})^{\star \bar k}\right]\ ,\label{calI}\eea
where one may take $\s=0$ if $k+\bar k>0$. For the perturbative regularization of the charges 
\bea {\cal I}(k+1,k,k)~=~\widehat{\rm Tr}'\left[\widehat\kappa\widehat{\bar\kappa}\star \widehat X^{\star k}\right]\ ,\eea
see \cite{Colombo:2010fu}, and for their evaluation on exact solutions, see \cite{Sezgin:2005pv,Iazeolla:2007wt}. 

Since the zero-form charges are closed on shell, one may consider $\widehat\Theta$-deformations in the case of manifestly Lorentz covariant models (and negative cosmological constant) of the form $\widehat\eta=\exp_\star (i\widehat \Theta)\star \widehat\Phi$ with $\widehat\Theta=\sum_{n=0}^\infty\widehat \Theta^{(2n)}$ where
\be
\widehat\Theta^{(2n)}~=~\sum_{\ell=0}^{\frac{n-1}2}
\sum_{\{k_i\}} \theta^{(2n)}_{\{k_i\}}\widehat X^{\star k_0}\prod_{i=1}^\ell {\cal I}(k_i+1,k_i,k_i) \ ,\ee
where $\widehat X:=\widehat \Phi\star \pi(\widehat \Phi)$, $\theta^{(2n)}_{\{k_i\}}\in \Real$ and $\{k_i\}$ with $i=0,\dots, \ell$ obey $\sum_{i=0}^\ell k_i = n$. It follows that $\widehat\Theta^{(2n)}$ is of $O(\widehat \Phi^{2n})$ and contains $2n-1$ integrals over ${\cal Y}\times {\cal Z}$. 
The original interactions, which contain no traces, correspond to taking $\theta^{(2n)}_{\{k_i\}}=0$ when $\ell>0$. 
The new interactions are as nonlocal in twistor space as the original ones.
While $\theta^{(2n)}_{\{k_i\}}$ are constants off shell, the functionals ${\cal I}(k_i+1,k_i,k_i)$ are constants only on shell. 
Relying on the results from \cite{Sezgin:2005pv,Iazeolla:2007wt,Colombo:2010fu}, the new interactions do not spoil the known exact solutions as the latter have well-defined ${\cal I}(k+1,k,k)$. 
Finally, we recall that adding the requirement of parity invariance implies that the $\widehat\Theta$-deformation can be redefined away, leaving only the Type A and Type B models described earlier.


\section{Geometrical Phase Based on $\widehat{\mathfrak{hs}}_+(4)\oplus {\mathfrak sl}(2,\Comp)$}\label{sec:5}

In this section we exhibit various observables and the soldering mechanism of the geometrical formulation of Vasiliev's theory based on the structure algebra $\widehat{\mathfrak{hs}}_+(4)$ given in \eq{hsplus}.

\subsection{$\pi$-Even Structure Group}

Taking the structure algebra to be the direct sum of the canonical Lorentz algebra and the $\pi$-even subalgebra $\widehat{\mathfrak{hs}}_+(4)$ of the full Cartan gauge algebra $\widehat{\mathfrak{hs}}(4)$ leads to connection $\widehat \C$ and soldering one-form $\widehat E$ given by
\be \widehat \C~=~ \widehat W^+\oplus  \o\ ,\qquad \widehat E~=~\widehat W^-\ ,\qquad \widehat W^\pm~=~\frac12 (1\pm\pi)\widehat W\ .\ee
Letting $\widehat \rho$ denote the representation of $\widehat{\mathfrak t}$ given by the direct sum of the adjoint representation of $\frac12 (1+\pi)\widehat{\mathfrak hs}(4)$ and the canonical representation $\rho$ of $\msl(2,\Comp)$ defined in \eq{LT1} and \eq{LT2}, we have the covariant derivative
\bea \widehat \nabla(\widehat \C)&=& d+\widehat\rho(\widehat\C)~=~\nabla+{\rm ad}_{\widehat W^+}\ ,\eea
where the Lorentz covariant derivative $\nabla$ is defined in \eq{nabla}.
Splitting also $\widehat \Phi$ and $\widehat S_\a$ into even and odd parts,
\be \widehat \Phi^\pm ~=~\frac12(1\pm \pi)\widehat \Phi\ ,\qquad
\widehat S_\a^\pm ~=~\frac12(1\pm \pi)\widehat S_\a\ ,\qquad
\widehat S_{\ad}^\pm ~=~ \frac12(1\pm \pi)\widehat S_{\ad}\ ,\ee
the master field equations in $X$-space can be written as
\be  \widehat R+\widehat E\star\widehat E+\widehat r^{(S)+}~=~0\ ,\qquad \widehat \nabla \widehat E+\widehat r^{(S)-}~=~0\ ,\label{xspace1}\ee\be \widehat\nabla \widehat \Phi^\pm+\left\{\widehat E,\widehat \Phi^\mp\right\}_\star~=~0\ ,\qquad\widehat \nabla\widehat S^\pm_\a+\left[\widehat E,\widehat S^\mp_\a\right]_\star~=~0\ ,\label{xspace2}\ee
where we have the curvature two-form
\be \widehat R ~=~\nabla\widehat W^++\widehat W^+\star\widehat W^++ \widehat r^{(0)}\ ,\ee
with Riemann two-form $\widehat r^{(0)}$ defined in \eq{widehatr},. We have also defined the projections
\be \widehat r^{(S)\pm}~=~\frac12(1\pm \pi)\widehat r^{(S)}\ ,\ee
of the quantity $\widehat r^{(S)}$ given in \eq{widehatr} using $\frac12(1\pm \pi)\widehat M^{(S)}_{\a\b}=\widehat S^+_\a\star\widehat S^\pm_\b+\widehat S^-_\a\star\widehat S^\mp_\b$. The Cartan integrability in particular makes use of
\be \widehat\nabla \widehat r^{(S)\pm}=-[\widehat E,\widehat r^{(S)\mp}]_\star\ .\ee
The locally defined gauge symmetries are
\begin{itemize}
\item the shift symmetry \eq{shiftsymm}; and
\item the generalized Lorentz transformations $\delta_{\widehat \L}\equiv\widehat\rho(\widehat \L)$ with parameter \be \widehat \L~=~\widehat\e^+\oplus \L\ ,\qquad \widehat \e^\pm~=~\frac12 (1\pm\pi)\widehat \e\ ,\ee
where the parameter $\L$ of the manifest local Lorentz symmetry acts in accordance with \eq{LT1} and \eq{LT2}, and $\widehat \e^+$ generates the transformations
\be \delta_{\widehat \e^+}\widehat W^+~=~ \widehat \nabla \widehat \e^+\ ,\qquad \delta_{\widehat\e^+}\o~=~0\ ,\qquad
\delta_{\widehat \e^+}\widehat E~=~ -[\widehat \e^+,\widehat E]_\star\ ,\ee\be
\delta_{\widehat \e^+}\widehat \Phi_\pm~=~ -[\widehat \e^+,\widehat \Phi_\pm]_\star\ ,\qquad
\delta_{\widehat \e^+}\widehat S^\pm_\a~=~ -[\widehat \e^+,\widehat S^\pm_\a]_\star\ .\ee
\end{itemize}
The broken gauge symmetries, referred to as generalized local translations, are
\be \delta_{\widehat \xi}\widehat \C~=~ [\widehat E,\widehat \xi]_\star\ ,\qquad \delta_{\widehat\xi}\o~=~0\ ,\qquad
\delta_{\widehat \xi}\widehat E~=~ \widehat\nabla \widehat \xi \ ,\ee\be
\delta_{\widehat \xi}\widehat \Phi_\pm~=~ -\{\widehat \xi,\widehat \Phi_\mp\}_\star\ ,\qquad
\delta_{\widehat \xi}\widehat S^\pm_\a~=~ -[\widehat \xi,\widehat S^\mp_\a]_\star\ ,\ee
where parameter is given by the $\pi$-odd projection
\bea \widehat \xi&=&\widehat \e^-\ .\eea
We recall that $(\widehat E,\widehat \xi)$ is a section of the $\widehat{\mathfrak{hs}}_+(4)\oplus {\mathfrak sl}(2,\Comp)$-bundle. Thus, on the overlap ${\underline{\cal M}}_I\cap{\underline{\cal M}}_{I'}$ the local representatives $(\widehat E_I,\widehat \xi_I)$ and $(\widehat E_{I'},\widehat \xi_{I'})$ are related by a transition function $\widehat G_I^{I'}$ generated by $\widehat{\mathfrak{hs}}_+(4)\oplus {\mathfrak sl}(2,\Comp)$. Moreover, the Lie derivative ${\cal L}_V$ along a globally defined vector field $V$ is equivalent on shell to the composite parameters
\be \widehat\e^+_V~=~i_V \widehat W^+\ ,\qquad {\L_V}~=~ i_V \o\ ,\qquad \widehat \xi_V\ =\ i_V\widehat E\ ,\qquad \eta^{\a\b}_V\ =\ \nabla\L^{\a\b}_V\ ,\label{Lie}\ee

\subsection{Abelian $p$-Form Charges}

The definition of a soldering one-form facilitates the construction of intrinsically defined classical observables given by the charges of on shell closed abelian $p$-form, \emph{viz.}
\bea  {\cal Q}[\Sigma,H]&=&\oint_{\S} H(\widehat E,\widehat \Phi,\widehat S_\a,\widehat S_{\ad}, r_{\a\b},\bar r_{\ad\bd})\ ,\label{calQ}\eea
where $\S$ are closed $p$-cycles in $\underline{\cal M}$ or open $p$-cycles with suitable boundary conditions, and $H$ are globally defined differential forms that are cohomologically nontrivial on shell, \emph{i.e.}
\be dH~=~0\ ,\qquad \delta_{\widehat \L}H~=~0 \qquad \mbox{off shell}\ ,\ee
and $H$ is not globally exact.
Thus, the charges ${\cal Q}$ are invariant
\begin{itemize}
\item off shell under generalized Lorentz gauge transformations (these are manifest symmetries of the charge densities);
\item on shell under diffeomorphisms in $\underline{\cal M}$ (modulo the equations of motion, the Lie derivative ${\cal L}_V H=d(i_V H)$ where $i_V H$ is globally defined implying that $\oint_{\S}d(i_V H)=0$);
\item on shell under deformations of $\S$ (which is to say that ${\cal Q}$ are generalized conserved charges).
\end{itemize}
The systematic search for abelian charges is tantamount to looking for cohomology groups in de Rham chain complexes consisting of globally defined differential forms. In particular, focusing on single chiral traces, one has complexes labelled by
\be (\overrightarrow{q};p)~\in~{\mathbb N}^7\ ,\qquad \overrightarrow{q}~=~(k,\tilde k;m,\tilde m;\bar m,\tilde{\bar m})\in {\mathbb N}^6\ ,
\ee
consisting of elements of the form
\be  M(\overrightarrow{q},p;\widehat K)~=~\widehat{\rm Tr}' \left[ \widehat M(\overrightarrow{q};p)\star \widehat K\right]\ ,\qquad \widehat K~\in~\left\{1,\widehat{\kappa},\widehat{\bar\kappa},\widehat\kappa\widehat{\bar\kappa}\right\}\ ,\ee
where $\widehat M(\overrightarrow{q};p)$ are $\star$-monomials of degree $\overrightarrow{q}$ in $(\widehat\Phi,\pi(\widehat \Phi);\widehat S_\a,\pi(\widehat S_\a);\widehat S_{\ad},\pi(\widehat S_{\ad}))\,$ and of total homogeneous degree $p$ in $(\widehat E,\widehat r^{(S)},\pi(\widehat r^{(S)}))$ where $\widehat r^{(S)}$ is defined in \eq{widehatr}.
In other words, $\widehat M(\overrightarrow{q};p)$ belongs to the linear space of multi-linear functions with the above degree assignments; these are generated by all possible permutations modulo the graded cyclicity and $\pi$-invariance of $\widehat Tr\left[(\cdots)\star\widehat K\right]$.

In the $\overrightarrow q=\overrightarrow 0$ sector, one has $\widehat {\rm Tr}'\left[ \widehat E^{\star p}\star \widehat K\right]\equiv 0$ for odd $p$ and all $\widehat K$ as well as for even $p$ and $\widehat K\in \left\{1,\widehat\kappa\widehat{\bar\kappa}\right\}$, using the parity arguments described above. Thus, for $\o=0$ the cohomology consists of $\widehat{\rm Tr}'\left[\widehat E^{\star p}\star \widehat \kappa\right]$ with $p=2,4,6,\dots$ and their hermitian conjugates. For finite $\o$, their $\widehat r^{(S)}$-dressed versions read ($p=2,4,6,\dots$)\,: \be  H_{[p]}~=~\sum_{r=0}^{p/2} { p \choose r} \widehat {\rm Tr}' \left[ \left\{\widehat E^{\otimes(p-2r)},(\widehat r^{(S)+})^{\otimes r}\right\}_{\rm symm}\star \widehat \kappa\right]\ ,\ee
where $\{\widehat X_1,\dots,\widehat X_N\}_{\rm symm}=\frac1{N!}\sum_{\s\in S_N} \widehat X_{\s(1)}\star\cdots \widehat X_{\s(N)}$ denotes the totally symmetric $\star$-monomial of degree $N$, that is
\be  H_{[p]}~=~ \widehat {\rm Tr}' \left[ (\widehat E\star\widehat E+\widehat r^{(S)+})^{\star (p/2)}\star \widehat \kappa\right]\ ,\label{Hp}\ee
whose conservation on shell can be checked directly using the fact that $\widehat\nabla(\widehat E\star\widehat E+\widehat r^{(S)+})=0$ on shell.

In sectors with nontrivial $\overrightarrow q$, we have examined various low-lying levels without finding any nontrivial cohomologies.

\subsection{Soldering Mechanism and Higher Tensorial Coordinates}

The nontriviality of the abelian $p$-form charges in \eq{calQ} requires $\Sigma$ to be a nontrivial closed $p$-cycle and  the soldering one-form $\widehat E$ to have at least rank $p$. However, the study of minimal areas and dynamical $p$-branes requires a soldering mechanism whereby the coset $\widehat{\mathfrak{hs}}(4)/\widehat{\mathfrak{hs}}_+(4)$ in the fiber, or a subspace thereof, becomes identified with the tangent space of ${\underline{\cal M}}$. The one-form $\widehat E$ then gives rise to a frame field $\widehat E_{\underline M}{}^{\underline A}$, also referred to as generalized vielbein, defined by
\be \widehat E_{\underline M}~=~\widehat E_{\underline M}{}^{{\underline A}} \widehat P_{ {\underline A}}\ ,\qquad \pi(\widehat P_{\underline A})~=~-\widehat P_{\underline A}\ ,\ee
where $\widehat P_{\underline A}$ are the coset generators. The local translations with gauge parameters $\widehat \xi=\widehat \xi^{\underline A} \widehat P_{{\underline A}}$ are then identified via \eq{Lie} as infinitesimal diffeomorphisms with globally defined vector field $\xi^{\underline M}=\widehat E^{\underline M}_{\underline A} \widehat \xi^{\underline A}$ combined with local generalized Lorentz transformations.

The soldering can be examined perturbatively by projecting the quasi-free differential algebra on ${\cal B}$ to a free differential algebra on a submanifold
\be {\cal M}~\subset~ \underline{\cal M}\ ,\ee
with local coordinates $X^M$. To this end, one first solves the constraints on ${\cal Z}$ given initial data in the form of the reduced master fields
\be W(X,Y)~\equiv ~dX^M W_M(X,Y)~=~\left.dX^{\underline M}\widehat W_{\underline M}\right|_{Z=0,{\cal M}}\ ,\qquad \Phi(X,Y)~=~\left.\widehat \Phi\right|_{Z=0}\ ,\ee
belonging to the reduced higher spin algebra $\mathfrak{hs}(4)$ and its twisted-adjoint representation $T[\mathfrak{hs}(4)]$ given by the $Z=0$ projections of \eq{hs4} and \eq{ths4}, respectively. Decomposing under $\mathfrak{hs}_+(4)$ and defining $\Gamma=\widehat \Gamma|_{Z=0}$, the remaining constraints on ${\underline{\cal M}}\times \{Z=0\}$ read
\be \nabla\Phi+[\C, \Phi]_\star+\{E, \Phi\}_\star+{\cal P}~=~0\ ,\ee\be  \nabla \C+\C\star\C+ E\star  E+{\cal J}^++r^+~=~0\ ,\qquad  \nabla E+[ \C, E]_\star+{\cal J}^-+r^-~=~0\ ,\ee
where ${\cal J}$ and ${\cal P}$ are quadratic and linear in $W$ and of $O(\Phi)$ and $O(\Phi^2)$, respectively, and $r$ is linear in $(r^{\a\b},\bar r^{\ad\bd})$ and of $O(\Phi^2)$. Projecting down to ${\cal M}$ using
\be E~=~\left.\left.\widehat E\right|_{Z=0}\right|_{{\cal M}}~=~ dX^M E_M^A P_A\ ,\qquad \pi(P_A)~=~-P_A\ ,\ee
where $P_{A}$ is a basis for the $\pi$-odd elements of ${\mathfrak{hs}}(4)$ and
$E_M^A$ is invertible, yields
\be \nabla_A(\o,\C)\Phi+\{P_A, \Phi\}_\star+{\cal P}_A=0\ ,\ee
\be R_{AB}+ P_A\star  P_B+{\cal J}^+_{AB}+r^+_{AB}~=~0\ ,\qquad T_{AB}+{\cal J}^-_{AB}+r^-_{AB}~=~0\ ,\ee
where we have defined covariant derivative $\nabla(\C,\o)=\nabla+{\rm ad}_\C$, curvature $R=\nabla \C+\C\star\C$ and generalized torsion $T=\nabla E+\left\{\C,E\right\}_\star$ and expanded in the frame field as follows:
\be \nabla(\C,\o)~=~E^A\nabla_A(\C,\o)\ ,\qquad R~=~\frac12 E^A E^B R_{AB}\ ,\qquad T~=~\frac12 E^A E^B T_{AB}\ .\ee
The resulting generalized $\sigma^-$-cohomology can be analyzed by decomposing $P_{A}=\left\{ P_{A_\ell}\right\}_{\ell=0}^\infty$ into levels of increasing tensorial rank, \emph{viz.}
\be P_{A_\ell}~=~\left\{P_{a(2\ell+1),b(2k)}\right\}_{k=0}^\ell~=~\left\{M_{\{a_1b_1}\star \cdots \star M_{a_{2k} b_{2k}}\star P_{a_{2k+1}}\star \cdots \star P_{a_{2\ell+1}\}}\right\}_{k=0}^{\ell}\ ,\ee
where $(M_{ab},P_a)$ are the generators of $\mathfrak{so}(3,2)$ and $P_{a(2\ell+1),b(2k)}$ is a Lorentz tensor of type $(2\ell+1,2k)$. The invertibility can be achieved by working in a triangular gauge for $E_M{}^A$ which requires the notion of a level expansion of the coordinates as well, say $X^M=\left\{ X^{M_\ell}\right\}_{\ell=0}^\infty$ where $X^{M_\ell}=\left\{X^{\mu(2\ell+1),\nu(2k)}\right\}_{k=0}^{\ell}$, such that it can be assumed that
\be \left.E_{M_\ell}{}^{A_{\ell'}}\right|_{X^{M_{\ell''}}=0}~=~0\ ,\qquad \ell>\ell'\ ,\quad \ell''>0\ .\label{bigE}\ee
The gauge function $L=\exp_\star (i X^M\delta_M^A P_A)$ leads to a formal definition of a higher spin generalization of $AdS_4$.
Its frame field and connection are contained in $L^{-1}\star dL$ which is given by non-Gaussian integrals for $X^{M_{\ell''}}\neq0$.
The latter can be expanded in $X^{M_{\ell''}}$ around $X^{M_{\ell''}}=0$ in terms of standard Gaussian integrals and hence \eq{bigE} is well-defined.
Whether $L$ is integrable for finite $X^{M_{\ell''}}$ remains to be studied.

The zeroth level of the zero-form constraint reads
\be \nabla_a(\C, \o)\Phi+\{P_a,\Phi\}_\star+{\cal P}_a~=~0\ ,\ee
where the translation generator is given by the twistor relation
\be P_a~=~\frac14 (\s_a)^{\a\ad}y_\a \yb_{\ad}\ ,\ee
which implies that $\Phi_{\a(n+2s),\ad(n)}$ is identifiable as the $n$th order symmetrized covariant vectorial derivative of the primary spin-$s$ Weyl tensor $\Phi_{\a(2s)}$ \cite{Us:analysis}. On the other hand, the $\ell$th level of the zero-form constraint implies\be \nabla_{a(2\ell+1)}(\C, \o)\Phi+\{P_{a(2\ell+1)},\Phi\}_\star~=~0\ ,\ee
where the higher translation generator is now given by the enveloping formula
\be P_{a(2\ell+1)}~=~P_{\{a_1}\star\cdots P_{a_{2\ell+1}\}}\ .\ee
As a result, the tensorial derivatives $\nabla_{a(2\ell+1)}(\C, \o)$ factorize into multiple vectorial derivatives; for example, the tensorial derivative of the physical scalar $\phi=\Phi|_{Y=0}$ factorizes into
\be \nabla_{a(2\ell+1)}(\C, \o)\phi~\propto~ \nabla_{\{a_1}(\C, \o)\cdots\nabla_{a_{2\ell+1}\}}(\C, \o)\phi\ .\ee
More generally, the derivatives $\nabla_{A_\ell}(\overline \C,\overline \o)C_{\a(2s)}$ are given by vectorial derivatives of primary Weyl tensors $\Phi_{\a(2s')}$ of spins $s'$ depending on $s$ and symmetry property of the tensorial coordinate $A_\ell$. For example,
\be \nabla_{a(2\ell+1),b(2k)}(\C,\o)\Phi~\propto~\nabla_{a(2\ell+1-2k)}(\C,\o)C_{a(2k),b(2k)}\ .\ee

Thus, for fixed $\ell$, the second order operators $\eta^{A(2\ell+1) B(2\ell+1)}\nabla_{A(2\ell+1)}(\C, \o)\nabla_{B(2\ell+1)}(\C, \o)$ are constrained: for $\ell=0$ the resulting second-order equations yield the well-known physical spectrum upon imposing suitable boundary conditions; for $\ell>0$ the resulting higher-derivative equations hold locally in a trivial fashion in view of the $\ell=0$ equations and the constraints. Not surprisingly, the nontrivial effects from the extension thus reside in boundary conditions; for example, a closed cycle of dimension $p$ with $p=6,8,\dots$ may activate the corresponding abelian $p$-form charge.

\subsection{Generalized Metrics and Minimal Areas}

Given a topologically nontrivial $p$-cycle $[\S]$ in $X$-space, one can construct classical observables as minimal areas with respect to rank-$s$ metrics
\be d\Sigma^s~=~dX^{{\underline M}_1}\cdots dX^{{\underline M}_s} G_{{\underline M}_1\dots {\underline M}_s}~=~ \eta_{(s)}(\widehat \Phi,\widehat S_\a,\widehat S_{\ad};\widehat E,\dots,\widehat E)\ ,\ee
where $\eta_{(s)}(\widehat \Phi,\widehat S_\a,\widehat S_{\ad};\cdot,\dots,\cdot)$ denotes $s$-linear and totally symmetric $\widehat {\mathfrak{t}}$-invariant functions\footnote{For an analog in three dimensional higher spin gravity, see \cite{Campoleoni:2010zq}.}. In other words, letting ${\cal A}[\Sigma',G_{(s)}]$ denote the area of the surface $\Sigma'\in [\Sigma]$, the minimal area \be {\cal A}_{\rm min}[\Sigma,G_{(s)}]~=~ \min_{\Sigma'\in [\Sigma]} {\cal A}[\Sigma',G_{(s)}]\ ,\ee
is an intrinsically defined and $\widehat t$-invariant classical observable.
A class of metrics are given by the totally symmetric parts of single chiral traces of strings of $\star$-products of $s$ generalized vielbeins $\widehat E=dX^{\underline M} \widehat E_{\underline M}$ with insertions of $s$ vertices ${\cal V}^{k_i,\bar k_i}_{\l_i,\bar \l_i}(\widehat \Psi,\widehat{\bar\Psi},\widehat S)$ ($i=1,\dots,s$) as given in \eq{vertex}, \emph{viz.}\footnote{Although we are here considering only totally symmetric metrics, one may in general consider metrics of more general symmetries.}
\be G_{{\underline M}_1\dots {\underline M}_s}~=~T_{(\underline M_1,\dots,\underline M_s)}\ ,\ee
where the rank-$s$ tensor
\be T_{\underline M_1,\dots,\underline M_s}~=~{\rm Tr}'\left[\widehat K\star \widehat E_{{\underline M}_1}\star {\cal V}^{k_1,\bar k_1}_{\l_1,\bar \l_1}\star \cdots\star \widehat E_{{\underline M}_s}\star {\cal V}^{k_s,\bar k_s}_{\l_s,\bar \l_s}\right]\ ,\ee
with $\widehat K\in\left\{1,\widehat{\kappa},\widehat{\bar\kappa},\widehat\kappa\widehat{\bar\kappa}\right\}$. Given such a metric, one introduces local coordinates on $\S'$, say $\s^m$, and computes the induced rank-$s$ metric
\be (f^\ast G)_{m_1\dots m_s}~=~\partial_{m_1} X^{{\underline M}_1}\cdots\partial_{m_s} X^{{\underline M}_s} G_{{\underline M}_1\dots {\underline M}_s}\ .\ee
Letting $\e^{m_1\dots m_p}$ denote the totally anti-symmetric tensor density on $\Sigma'$, one can then take
\be  {\cal A}[\S',G_{(s)}]~=~\int_{\S'} d^{p}\s \left(I\left(\e^{\otimes k},(f^\ast G_{(s)})^{\otimes l}\right)\right)^{1/k}\ ,\qquad pk~=~sl\ ,\ee
where $I$ is a scalar density of degree $k$. For example, if $s$ is even one may take the Cayley-style determinants
\be I~=~\e^{m_1[p]}\cdots \e^{m_s[p]} \underbrace{(f^\ast G)_{m_{1}\dots m_{s}}\cdots (f^\ast G)_{m_{1}\dots m_{s}}}_{\tiny \mbox{$p$ times}}\ ,\qquad k=s\ ,\quad l=p\ ,\ee
whereas odd $s$ require more involved invariants. In the case of $s=2$ the Riemann curvature $R_{M(2);N(2)}$ (symmetric convention) has a perturbative expansion in terms of generalized Riemann tensors, \emph{viz.} $R_{\mu,\mu(s-1);\nu(1),\nu(s-1)}\sim C_{\m(s),\n(s)}$. The tensorial calculus pertinent to examining the variational principles for $s>2$ remains to be investigated further.

There are infinitely many inequivalent metrics.
Whether there is any principle to single out any preferred metric $G_{\underline M_1\dots\underline M_s}=T_{(M_1,\dots,M_s)}$ remains to be studied.
To this end, one may speculate that it is possible to derive Vasiliev's equations from a suitable consistency requirement for a dynamical $p$-brane with action given by the $p$-volume and possibly couplings to anti-symmetric forms and other poly-forms given by various Young projections of $T_{M_1,\dots,M_s}$.
One may also envisage the emergence of a preferred metric within the framework of a higher-spin generalization of the superembedding approach to superbranes, possibly in the context of infinite-dimensional geometries based on the coset $\mathfrak{hs}(4)/\mathfrak{hs}_+(4)$. 

\section{On shell Deformations of Bulk Action and Tree Amplitudes}

In this section we shall construct off shell boundary deformations of the recently proposed action for Vasiliev's four dimensional higher spin gravity given in \cite{Boulanger:2011dd}
that reduce on shell to the zero-forms \eq{calI} and the $p$-forms \eq{Hp}.
The off shell deformations, that we shall refer to as topological defects, or topological vertex operators, are invariant off shell under gauge transformations forming a structure group.
Moreover, they do not contribute to the classical equations of motion and preserve the on shell boundary conditions.
Thus, in the semi-classical limit, their effect is to produce a nontrivial contribution to the action on shell that can be interpreted as a classical amplitude.

Bulk actions for the minimal bosonic models can be obtained by consistent truncation of enlarged models with external Kleinian operators $(K,\bar K)$ \cite{Boulanger:2011dd}; the enlargement amounts to the additional $\star$-product rules
\be \{K,y^\a\}_\star~=~\{K,z^\a\}_\star~=~\{K,dz^\a\}_\star~=~0\ ,\qquad K\star K~=~1\ ,\ee
\emph{idem} $\bar K$ and $(\yb^{\ad},\zb^{\ad},d\zb^{\ad})$.
In particular, an action with linear $\eta_1$ and quadratic Poisson structure is given by
\be S_{\rm bulk}~=~\int_{{\cal M}} \widehat {\rm Tr}\left.\left[\frac{1+K\bar K}2\star \left(\widehat U\star \widehat D\widehat B+\widehat V\star(\widehat F+\widehat B\star \widehat J+\widehat U)\right)\right]\right|_{K=\bar K=0}\ ,\ee
where ${\cal M}$ is odd-dimensional; $\widehat J=\widehat J_{[2]}+\widehat J_{[4]}$ is closed and central term; $(\widehat A,\widehat V)$ have odd form degree and $(\widehat B,\widehat U)$ have even form degree; and the evaluation at $K=\bar K=0$ is to be carried out after performing the $\star$-products.

For example, if ${\rm dim}{\cal M}=5$, then
\be \widehat B ~=~ \widehat B_{[0]}+\widehat B_{[2]}+\widehat B_{[4]}+\widehat B_{[6]}\ ,\qquad \widehat A ~=~ \widehat A_{[1]}+\widehat A_{[3]}+\widehat A_{[5]}+\widehat A_{[7]}\ ,\ee
\be \widehat U ~=~ \widehat U_{[2]}+\widehat U_{[4]}+\widehat U_{[6]}+\widehat U_{8]}\ ,\qquad\widehat V ~=~ \widehat V_{[1]}+\widehat V_{[3]}+\widehat V_{[5]}+\widehat V_{[7]}\ .\ee
The variational principle implies the Cartan integrable bulk equations of motion
\be \widehat D\widehat B+\widehat V~\approx~0\ ,\qquad \widehat F+\widehat B\star\widehat J+\widehat U~\approx~0\ ,\ee
\be \widehat D\widehat U -\widehat V\star\widehat J~\approx~0\ ,\qquad \widehat D\widehat V+\left[\widehat B,\widehat U\right]_\star~\approx~0\ , \ee and $\int_{\partial{\cal M}} \widehat {\rm Tr}[\widehat U\star\delta \widehat B-\widehat V\star\delta\widehat A]\approx 0$. The latter imply
\be \widehat U|_{\partial {\cal M}}~=~0\ ,\qquad \widehat V|_{\partial {\cal M}}~=~0\ ,\ee
which one may view as part of the definition of a generalized Hamiltonian action principle.
The on shell Cartan gauge transformations are symmetries of the action as follows: the gauge parameters $\widehat\e$ associated with $(\widehat A,\widehat B)$ generate transformations that leave the Lagrangian invariant, while those of $(\widehat U,\widehat V)$ do so up to total derivatives and must hence form a section. Taking into account the boundary conditions, it thus follows that $(\widehat U,\widehat V)$ can be set equal to the zero section on shell by fixing a gauge, leaving
\be \widehat F+\widehat B\star\widehat J~=~0\ ,\qquad \widehat D\widehat B~=~0\ ,\ee
which is a duality extended version of Vasiliev's original system of equations with linear interaction freedom. Assuming that the gauge symmetries associated with the master fields of form degrees $p\geq 2$ remain unbroken, the duality extended and original systems share the same initial and boundary data, and are hence equivalent on shell (modulo transition functions) as discussed in Section \ref{sec:data}.

Since the bulk action vanishes on shell, semi-classical amplitudes can only be generated by deformations $S_{\rm top}$, referred to as topological vertex operators, which are functionals integrated over submanifolds of ${\cal M}$ where the Lagrange multipliers $(\widehat U,\widehat V)$ vanish, with the following properties:
\begin{itemize}
\item $\delta_{\widehat\e} S_{\rm top}=0$ for unbroken gauge parameters $\widehat \e$ associated with $(\widehat A,\widehat B)$;
\item the general variation $\delta S_{\rm top}\approx 0$ on the shell of $S_{\rm bulk}$ which implies that the equations of motion as well as the boundary conditions of the deformed theory remain the same as those of the undeformed theory; and
\item $S_{\rm top}$ has a nontrivial value on shell.
\end{itemize}

One simple class of off shell deformations consists of the following generating functionals for twistor space amplitudes in the unbroken phase ($n=0,1,2,\dots$ ; $\widehat K\in\{1,k\k,\bar k\bar\k,k\bar k\k\bar \k\}$):
\be S^{\rm top}_{n,\widehat K}~=~\left.\left.\widehat{\rm Tr}\left[\frac{1+K\bar K}2\star \widehat K\star \left(\widehat F\star \widehat B^{\star n}+\frac{n}{n+1}\widehat B^{\star (n+1)}\star \widehat J\right)\right]\right|_{K=\bar K=0}\right|_{p}\ ,\label{stop}\ee
\be S^{{\rm top}\prime}_{n,\widehat K}~=~\left.\left.\widehat{\rm Tr}\left[\frac{1+K\bar K}2\star \widehat K\star \left(\widehat F^{\star 2}\star \widehat B^{\star n}-\frac{n}{n+2}\widehat B^{\star (n+2)}\star \widehat J^{\star 2} \right)\right]\right|_{K=\bar K=0}\right|_{p}\ ,\label{stopprime}\ee
where the point $p\in {\cal M}$ is chosen such that
\be \widehat U|_{p}~=~0\ ,\qquad \widehat V|_{p}~=~0\ .\ee
Their on shell values are given by
\be S^{\rm top}_{n,\widehat K}~\approx~-\frac1{n+1}\left.\widehat{\rm Tr}\left[\frac{1+K\bar K}2\star\widehat K\star \widehat B^{\star(n+1)}\star \widehat J\right]\right|_p\ ,\ee
\be S^{{\rm top}\prime}_{n,\widehat K}~\approx~\frac2{n+2}\left.\widehat{\rm Tr}\left[\frac{1+K\bar K}2\star \widehat K\star \widehat B^{\star (n+2)}\star \widehat J^{\star 2} \right]\right|_p\ .\ee
In order to make contact with the zero-form charges in \eq{calI}, we truncate the above fields to those of the duality extended minimal bosonic models given by \cite{Boulanger:2011dd} as follows:
\be \widehat A_{[1]}~=~\frac{1+K \bar K}2\star \widehat A_{[1]}\ ,\qquad \widehat A_{[3]}~=~\widehat C_{[3]}\star \left(\frac{K+ \bar K}2\right)^{\eta} \ ,\ee
\be \widehat B_{[0]}~=~ \widehat \Phi\star \frac{1+K \bar K}2\ ,\qquad \widehat B_{[2]}~=~\widehat C_{[2]}\star \left(\frac{K+ \bar K}2\right)^{\eta+1} \ ,\ee
where $\eta=0,1$ (corresponding to $\e_k=+1,-1$ in \cite{Boulanger:2011dd} and we note that the dual fields do not appear in the equations of motion for the original master fields).
For suitable $\widehat K$, the deformations of type $S^{{\rm top}\prime}$ reproduce the zero-form invariants in \eq{calI}, \emph{viz.}
\be S^{{\rm top}\prime}_{2m,\widehat{K}}~\approx~\frac1{m+1}\left.\widehat{\rm Tr}'\left[\widehat{K}\star \widehat X^{\star(m+1)}\right]\right|_p\ ,\qquad \widehat{K}~=~1\,,\, \widehat\k\widehat{\bar \k}\ ,\ee
\be S^{{\rm top}\prime}_{2m+1,\widehat{K}}~\approx~\frac 2{2m+1}\left.\widehat{\rm Tr}'\left[\widehat{K}\star \widehat X^{\star m}\star \widehat\Phi\right]\right|_p\ ,\qquad \widehat{K}~=~\widehat\k\,,\,\widehat{\bar\kappa}\ ,\ee
where $\widehat X:=\widehat\Phi\star\pi(\widehat \Phi)$.
The on shell values of the deformations of type $S^{\rm top}_{n,\widehat K}$, on the other hand, are given by linear combinations of \eq{calI} and additional, separately on shell closed, zero-form invariants involving $\widehat \Phi$ as well as $\widehat C_{[2]}$; see Appendix \ref{app:B}. In the latter case, the duality extension is nontrivial in the sense that it gives rise to classical observables that are functionals of the initial data $C(Y)=\left.\widehat \Phi\right|_{Z=0,p}$ that cannot be given a local description within the original duality unextended system.

Two examples of off shell deformations related to the $p$-forms \eq{Hp}, are the two-form\be S_{\rm top}[\S_2]~=~{\rm Re}\left\{ \tau_2\oint_{\S_2}
\widehat{\rm Tr}'\left[\widehat \kappa \star \widehat R\right]\right\}\ ,\ee
and the four-form
\be S_{\rm top}[\S_4]~=~{\rm Re}\left\{ \oint_{\S_4}
\widehat{\rm Tr}'\left[\widehat \kappa\star \left(  \tau_4 \widehat R\star \widehat R+ \tilde \tau_4\left((\widehat E\star\widehat E+\widehat r^{(S)+})\star \widehat R+\frac12 (\widehat E\star\widehat E+\widehat r^{(S)+})^{\star 2}\right)\right)\right]\right\}\ ,\label{tvo}\ee
where $\tau_2$, $\tau_4$ and $\tilde \tau_4$ are complex constants and $\S_{2,4}$ are submanifolds of ${\cal M}$ where the Lagrange multipliers vanish. The on shell value of the two-form is given by
\be S_{\rm top}[\S_2]~\approx ~ - {\rm Re}\left\{\tau_2 \oint_{\S_2}H_{[2]}\right\}\ ,\ee
with $H_{[2]}$ given in \eq{Hp}.
One application of this surface operator is to wrap $\Sigma_2$ around a point-like defect or singularity such as the center of the rotationally symmetric and static solution of \cite{Didenko:2009td}. In this case, the leading order contribution, which comes from the anti-de Sitter vacuum,  is a divergent integral over twistor space. If the divergence has a definite reality property, one can cancel it by choosing $\tau_2$ appropriately. One would then be left with an integral over the perturbations. As the latter involve nontrivial functions in twistor space, the integral may be finite. It would be interesting to seek an interpretation of the resulting value of $S_{\rm top}[\S_2]$ as some form of entropy of the Didenko -Vasiliev solution.

Turning to the four-form deformation, its on shell value reads

\be
S_{\rm top}[\S_4]~\approx~ {\rm Re}\left\{(\t_4-\frac12 \tilde \tau_4)\oint_{\S_4}H_{[4]}\right\}\ ,\label{ampl}\ee
where $H_{[4]}$ given in \eq{Hp}. Infinities now arise from the integration over $\Sigma_4$ as well as twistor space. Assuming once again that the divergence from the anti-de Sitter vacuum has a definite reality property, such that it can be removed by choosing $\tau_4$ appropriately, one is left with an integral over perturbations that may in principle be finite modulo a prescription for integration contours; for related discussions, see \cite{Giombi:2010vg,Colombo:2010fu}. In this case, and provided one considers perturbations corresponding to boundary sources, it would be natural to interpret the resulting $S_{\rm top}[\S_4]$ as the generating functional for the boundary correlation functions.

\section{Conclusions}

In this paper we have proposed geometric formulations of Vasiliev's higher spin gravity with generalized vielbeins and curvature zero-forms forming sections of bundles associated to a principle bundle of a structure group.
The choice of structure group is not unique.
In this paper, we have proposed four possibilities and considered two of them in detail,
namely the unbroken phase and a broken phase.
In the former, the structure algebra is the full higher spin algebra plus ${\mathfrak sl}(2,\Comp)$.
In the latter, it is defined by a projection that utilizes an automorphsim of the $\star$-product algebra.
In both cases we have constructed intrinsically defined classical observables by tracing the suitable constructs over the $\star$-product algebra.
The observables are manifestly invariant off shell under the structure group gauge transformations and invariant on shell under the remaining local higher spin translations.
In the unbroken phase, the observables are Wilson loops and certain zero-form charges, that we in addition used to extend the interactions of Vasiliev's theory.
In the broken phase, we have constructed two types of observables: generalized metrics and corresponding minimal area functionals and charges of on shell closed abelian $p$-forms.
We have shown that the latter have a natural interpretation as semi-classical amplitudes within the off shell formulation proposed in \cite{Boulanger:2011dd}.

The studies in this paper leave a number of issues open. In particular, it remains to be examined whether the four-form \eq{tvo} generates the standard correlation functions of the Vasiliev theory of which the two- and three-point functions are available in the work of Giombi and Yin \cite{Giombi:2010vg,Giombi:2009wh}.
It would also be interesting to compare the four-form to the Fradkin--Vasiliev cubic action \cite{Fradkin:1986qy}.
Moreover, the arguments presented above for how to regularize the on shell action indicate that the on shell value of the classical action vanishes for the anti-de Sitter vacuum.
As this value is related to the free energy of the dual conformal field theory, this raises a puzzle if one assumes that the latter is a free field theory.
As the deformation that we have proposed here is by means unique there may be a resolution to this puzzle by considering either further deformations.
In particular, it would be interesting to investigate alternative structure algebras such as the one given by the holomorphic enveloping algebra extension of the Lorentz algebra as described in this paper.

Finally, a study that may shed light on the first-quantized origin of Vasiliev's theory, is to compare the twistor space amplitudes arising from \eq{stop} and \eq{stopprime} with the topological open string in twistor space studied in \cite{Engquist:2005yt}.

\textbf{Acknowledgment}

We are grateful to N. Boulanger for collaborations at certain stages of this work. We thank C.~Iazeolla and M.~Vasiliev for illuminating discussions. We both thank Scuola Normale Superiore in Pisa for hospitality during early stages of this work. E.~S. thanks the University of Mons and P.~S. thanks the Mitchell Institute for Fundamental Physics and Astronomy for hospitality. The research of E.~S. is supported in part by NSF grants PHY-0555575 and PHY-0906222.


\begin{appendix}


\section{Lorentz Covariantization}\label{App:B}

The canonical Lorentz transformations and Lorentz covariant derivatives are defined by
\bea \delta_\L\psi_\a&=& \L_\a{}^\b \psi_\b\ ,\qquad \d_\L\o_M^{\a\b}\ =\
\partial_M \L^{\a\b}-2\o_M^{\c(\a}\L_\c{}^{\b)}\ ,\eea\be \nabla\psi_\a ~=~d\psi_\a -
\o_{\a}{}^\b\psi_\b\ ,\qquad R^{\a\b}~=~d\o^{\a\b}-\o^{\a\c}\o_\c{}^\b\ .\ee
The master field equations imply that the Cartan gauge transformations $\delta_{\widehat \e_\L}$ with parameters
\bea \widehat \e_\L&=& \frac1{4i} \left(\L^{\a\b} \widehat M_{\a\b}+\bar\L^{\ad\bd} \widehat M_{\ad\bd}\right)\ ,\eea
generate the canonical Lorentz transformations \eq{LT1} and \eq{LT2} of the full
Vasiliev system. To demonstrate this, one first notes that by their very definition, a Cartan gauge transformation $\delta_{\widehat\e}$ acts on a
general composite $\widehat X=\sum_{m,n} Y^m Z^n X_{m,n}$ where $X_{m,n}$ are functionals of the components of basic master fields, as $\delta_{\widehat\e}\widehat X=
\sum_{m,n}Y^m Z^n \delta_{\widehat \e} X_{m,n}$; in particular, one has $\delta_{\widehat \e} y_\a=0=\delta_{\widehat\e} z_\a$. To proceed, one shows that
\be[\widehat M_{\a\b},\widehat\Phi]_\pi\ =\ [\widehat M^{(0)}_{\a\b},\widehat \Phi]_\star\ ,\qquad [\widehat M_{\a\b},\widehat S_\c]_\star\ =\ [\widehat M^{(0)}_{\a\b},\widehat S_\c]_\star-4i\e_{\c(\a}\widehat S_{\b)}\ ,\label{lemmas}\ee
where the last equation follows from
\be [\widehat M^{(S)}_{\a\b},\widehat S_\c]_\star~=~ -4i\e_{\c(\a}\widehat S_{\b)}\ .\label{sss}\ee
It follows that the $\widehat S$-dependent terms drop out from
$\delta_\L \widehat \Phi$ and $\d_\L\widehat S_\a$, that hence assume the form given in \eq{LT1} and \eq{LT2}. From \eq{sss} it also follows that $\delta_\L\widehat A_\a= \L_\a{}^\b\widehat A_\b-[\widehat\e^{(0)}_\L,\widehat A_\a]_\star$ \emph{idem} $\widehat
A_{\ad}$. For $\widehat W_M$ one has
\bea \delta_\L \widehat W_M&=& \delta_\L \widehat A_M -\frac1{4i} \left(\delta_\L (\o_M^{\a\b} \widehat M_{\a\b})+\delta_\L (\bar
\o_M^{\ad\bd} \widehat M_{\ad\bd})\right)\ ,\eea
where $\delta_\L\o^{\a\b}$ is given above and
\be \delta_\L\widehat A_M~=~ \frac 1{4i}\left((\partial_M \L^{\a\b})\widehat M_{\a\b} +(\partial_M \bar\L^{\ad\bd})\widehat M_{\ad\bd}\right)
-[\widehat\e^{(0)}_\L,\widehat A_M]_\star\ ,\ee\\[-20pt]\be \delta_\L \widehat M_{\a\b}~=~ \delta_\L \widehat S_{(\a}\star \widehat S_{\b)}+\widehat S_{(\a}\star\delta_\L \widehat S_{\b)}~=~ 2\L_{(\a}{}^\c \widehat M_{\b)\c}-[\widehat\e^{(0)}_\L,\widehat M_{\a\b}]_\star\ ,\ee
implying that the terms containing $\partial_M \L^{\a\b}$ cancel such that $\widehat W_M$ transforms as in \eq{LT1}.

Furthermore, substituting the redefinition of $\widehat A_M$ into the constraints on $d \widehat \Phi$ and $d \widehat S_\a$, and using \eq{lemmas}, one finds \eq{cS}. One can the proceed amd substitute the redefinition into the constraint on $d\widehat A$; using \eq{cS} and \eq{sss} one can then reduce the quartic $\star$-product of $\widehat S_\a$'s down to a quadratic one after which \eq{cW} follows by the definition of $\nabla \widehat W$ and $R^{\a\b}$.

Finally, we note that the full Lorentz generators $\widehat M_{\a\b}$ form the field-dependent algebra
\bea [\widehat M_{\a\b},\widehat M_{\c\d}]_\star&=& 2i\e_{\b\c}\widehat M_{\a\d}+\mbox{$3$ perms}-
\d_{\L;\a\b}\widehat M_{\c\d}+\d_{\L;\c\d}\widehat M_{\a\b}\ ,\eea
which obeys the Jacobi identity. To this end, if $\mathfrak g$ is a Lie algebra with generators $X$ represented in an associative $\star$-product algebra spanned by $\widehat V^r$ by
\bea \rho(X)\ :=\ \widehat V^r \rho_r(X;\phi)\ ,\eea
where $\rho_r(X;\phi)$ are functions on a manifold with local coordinates $\phi^i$ carrying a representation of $\mathfrak g$ given by Lie derivatives ${\cal L}_X$, \emph{viz.}
\be {\cal L}_X \rho(Y) ~=~\widehat V^r {\cal L}_X \rho_r(Y;\phi)\ ,\qquad {\cal L}_X \rho_r(Y;\phi)~=~ X^i\partial_i \rho_r(Y;\phi)\ ,\ee
then it follows that the representation property
\be [\rho( X),\rho( Y)]_\star~:=~\rho([X,Y])-{\cal L}_X \rho(Y)+{\cal L}_Y\rho(X)\ ,\ee
is compatible with associativity and $[{\cal L}_X,{\cal L}_Y]={\cal L}_{[X,Y]}$.

\section{Additional Zero-Form Invariants in the Duality Extended Model }\label{app:B}

For suitable $\widehat K$, the on shell values of the deformations $S^{\rm top}$ in \eq{stop} are given for $\eta=0$ by
\be
S^{\rm top}_{2m,\s}~\approx~\frac{-1}{2m+1}\left.\widehat{\rm Tr}'\left[(\widehat\k\widehat{\bar \k})^\sigma\star \widehat X^{\star m}\star\left(\widehat\Phi \star \widehat J_{[4]} +(2m+1) \widehat C_{[2]}\star \widehat J_{[2]}\right)\right]\right|_p\ ,\ee
\be S^{\rm top}_{2m+1}~\approx~\frac{-1}{2(m+1)}\left.\widehat{\rm Tr}'\left[\widehat\k\star \widehat X^{\star m}\star\widehat\Phi\star\left(\pi(\widehat\Phi)\star \widehat J_{[4]} +2(m+1) \pi(\widehat C_{[2]})\star \widehat J_{[2]}\right)\right]\right|_p\ ,\qquad\ee
where $\sigma=0,1$, and for $\eta=1$ by
\be
S^{\rm top}_{2m}~\approx~\frac{-1}{2m+1}\left.\widehat{\rm Tr}'\left[\widehat\k\star \widehat X^{\star m}\star\left(\widehat\Phi \star \widehat J_{[4]} +(2m+1) \widehat C_{[2]}\star \widehat J_{[2]}\right)\right]\right|_p\ ,\ee\be
S^{\rm top}_{2m+1,\s}~\approx~\frac{-1}{2(m+1)}\left.\widehat{\rm Tr}'\left[(\widehat\k\widehat{\bar\k})^\s \star \widehat X^{\star m}\star \widehat\Phi\star \left(\pi(\widehat\Phi)\star \widehat J_{[4]} +2(m+1) \pi(\widehat C_{[2]})\star \widehat J_{[2]}\right)\right]\right|_p\ ,\qquad\qquad
\label{c1234}
\ee
which we identify as linear combinations of the zero-form invariants \eq{calI} of the duality unextended system and a new set of invariants involving $\widehat C_{[2]}$. The latter are closed on shell as can be seen using \eq{master2} and the additional duality extended equations of motion at the point $p$, which read
\be \widehat D \widehat C_{[2]}+ \left[\widehat C_{[3]},\widehat \Phi\right]_\pi~=~0\ ,\qquad  \widehat D \widehat C_{[3]}+\widehat \Phi \star \widehat J_{[4]}+\widehat C_{[2]}\star \widehat J_{[2]}~=~0\ ,\label{masternew}\ee
where the covariant derivatives and the projected $\widehat J_{[4]}$ are defined by
\bea \eta~=~0&:& \widehat D \widehat C_{[2]}~=~\widehat d \widehat C_{[2]}+\left[\widehat A_{[1]} ,\widehat C_{[2]}\right]_\pi \ ,\qquad \widehat D \widehat C_{[3]}~=~\widehat d\widehat C_{[3]}+\left\{ \widehat A_{[1]}, \widehat C_{[3]}\right\}_\star \ ,\\&& \widehat J_{[4]}~=~i (c \kappa+\bar c\bar\k)d^2z d^2\bar z\ ,\qquad c~\in~\Comp\ ,\\[5pt]
\eta~=~1&:& \widehat D \widehat C_{[2]}~=~\widehat d \widehat C_{[2]}+\left[\widehat A_{[1]}, \widehat C_{[2]}\right]_\star\ ,\qquad \widehat D \widehat C_{[3]}~=~\widehat d\widehat C_{[3]}+\left\{\widehat A_{[1]}, \widehat C_{[3]}\right\}_\pi\ ,\\ && \widehat J_{[4]}~=~ (\l +\l'\k\bar \k)d^2z d^2\zb\ , \qquad \l,\l'~\in~\Real\ .\eea

\end{appendix}

\newpage


\end{document}